\documentclass[preprint,superscriptaddress,nofootinbib,prd]{revtex4-1}
\usepackage[english]{babel}
\usepackage[utf8]{inputenc}

\usepackage{array}
\usepackage{graphicx}
\usepackage{epstopdf}

\usepackage{subfig}

\usepackage{enumerate}
\usepackage{dcolumn}

\usepackage[namelimits]{amsmath}
\usepackage{amssymb}
\usepackage{amsfonts}
\usepackage{bbm}      

\usepackage{hyperref}

\hypersetup{colorlinks=true, linkcolor=black, citecolor=black, filecolor=black, pagecolor=black, urlcolor=black}

\usepackage{multirow}
\usepackage{color}

\usepackage[T1]{fontenc}      
\usepackage{ae}               
\usepackage{indentfirst}
\usepackage{comment}

\usepackage[normalem]{ulem}  

\usepackage{tikz}
\usetikzlibrary{decorations.markings}

\newcommand{\sigmatot}{\sigma_{\text{tot}}}
\newcommand{\sigmael}{\sigma_{\text{el}}}

\DeclareMathOperator{\Imag}{Im}

\newcommand{\AP}{A_\mathbb{P}}   
\newcommand{\pp}{pp}
\newcommand{\ppbar}{\bar{p}p}

\newcommand{\LT}{\text{LT}}
\newcommand{\SLT}{\text{SLT}}
\newcommand{\LTth}{\text{LT}_\text{th}}
\newcommand{\SLTth}{\text{SLT}_\text{th}}

\newcommand{\SLTthk}{\text{SLT}_{\text{th},\kappa}}

\begin{document}
 
 
 \title{An investigation on the leading and subleading high-energy
   behavior of hadron-hadron total cross sections using a best-fit
   analysis of hadronic scattering data}
 
 \date{\today}

 \author{M. Giordano}
 \email{giordano@bodri.elte.hu}
 \affiliation{Institute for Theoretical Physics, E\"otv\"os University,
  and MTA-ELTE ``Lend\"ulet" Lattice Gauge Theory Research Group,
  P\'azm\'any P. s\'et\'any 1/A, H-1117 Budapest, Hungary}
  \author{E. Meggiolaro}
 \email{enrico.meggiolaro@unipi.it}
 \affiliation{Dipartimento di Fisica dell'Universit\`a di Pisa, and INFN,
  Sezione di Pisa, Largo Pontecorvo 3, I-56127 Pisa, Italy} 
 \author{P.V.R.G. Silva}
 \email{precchia@ifi.unicamp.br}
 \affiliation{Instituto de F\'isica Gleb Wataghin, Universidade
  Estadual de Campinas - UNICAMP, 13083-859 Campinas, SP, Brazil} 

 \begin{abstract}
 
 In the present investigation we study the leading and subleading
 high-energy behavior of hadron-hadron total cross sections using a
 best-fit analysis of hadronic scattering data. The parametrization
 used for the hadron-hadron total cross sections at high energy is
 inspired by recent results obtained by Giordano and Meggiolaro (2014)
 using a nonperturbative approach in the framework of QCD and it reads
 $\sigmatot \sim B\ln^2 s + C\ln s\ln\ln s$. Both $B$ and $C$ are
 obtained by means of best-fits to data for proton-proton and
 antiproton-proton scattering, including recent data obtained at the
 LHC, and also to data for other meson-baryon and baryon-baryon
 scattering processes. The results are compared to the theoretical
 predictions existing in the literature.
 In particular, following the above-mentioned nonperturbative QCD approach,
 we also consider fits where the parameters $B$ and $C$ are set to
 $B = \kappa B_\text{th}$ and $C = \kappa C_\text{th}$, where $B_\text{th}$
 and $C_\text{th}$ are universal quantities related to the QCD stable
 spectrum, while $\kappa$ (treated as an extra free parameter) is related
 to the asymptotic value of the ratio $\sigmael/\sigmatot$.
 Different possible scenarios are then considered and compared.

 \end{abstract}

  
 \maketitle

 

\section{Introduction}
\label{sec:Intro}

It is widely believed that the correct description of the strong
interaction between quarks and gluons is provided by Quantum
Chromodynamics (QCD), a belief supported by the numerous experimental
verifications obtained in more than forty years. There are, however,
energy regimes where the comparison between theory and experiment has
not been performed yet, due to the lack of a theoretical prediction
from the first principles of QCD. Such regimes are those where a major
role is played by the nonperturbative dynamical aspects of QCD, which
are notoriously very difficult to study. A successful comparison of
experiment and theory in one of these regimes would certainly provide
even stronger support to QCD being the appropriate description of 
strong interactions.

The total cross section ($\sigmatot$) for hadron-hadron
scattering processes at high energy is one of the best known
observables for which a fully satisfactory prediction from first principles
is not yet available. The reasons why the theoretical study
of this quantity in QCD is extremely difficult is better understood
recalling the relation between $\sigma_{\rm tot}$ and the forward
scattering amplitude provided by the optical theorem, which at high
energy reads
\begin{equation}
 \sigmatot(s) \mathop{=}_{s\to\infty} \frac{\Imag F(s,t=0)}{s}\,,
 \label{eq:optical_theorem}
\end{equation}
where $s$ is the total center-of-mass energy squared and
$t=-\vec{q}^{\,2}$ is the transferred momentum squared. The optical
theorem shows that 
the study of $\sigma_{\rm tot}$ requires the understanding of QCD in the
regime of very low momentum transfer where the strong coupling
constant becomes large~\cite{Halzen_Martin_book1984}, i.e., deep in
the nonperturbative regime. On the other hand, at large $s$ one cannot
employ the main tool used in nonperturbative investigations of QCD,
namely the lattice formulation of the theory, to compute directly the
relevant quantities. 
The most interesting features of $\sigmatot$, which should be
explained by QCD, are its increase with energy at large energy, and
the fact that the leading term is apparently universal, i.e.,
independent of the type of particles initiating the process.

More generally, the same limitations discussed above apply to the
study of soft high-energy hadron-hadron scattering, i.e., scattering
at large $s$ and small $|t|\lesssim 1~\text{GeV}^2$. Consequently, a full,
model-independent description, obtained from the first principles of
QCD, is not yet available for processes such as forward and
near-forward elastic scattering. The study of soft high-energy physics
has so far mostly proceeded through phenomenological models (for a
recent review see Ref.~\cite{Pancheri_Srivastava:2016}), on one
side, and through general results obtained using basic properties of
the theory, like unitarity and analyticity. Such general results are
usually incorporated in the phenomenological models, for example
exploiting them to set constraints on the functional form of
observables concerning their dependence on energy and 
momentum transfer. Regarding hadronic total cross sections,
the most important such result is probably the Froissart-\L ukaszuk-Martin (FLM)
bound~\cite{Froissart1961,Martin1963,Martin1965}, which states that $\sigmatot$
cannot grow asymptotically faster than $\ln^2 s$, i.e.,
\begin{equation}
 \sigmatot(s) \leq B_\text{FLM} \ln^2 (s/s_0) \quad (s\to\infty)\,,
 \label{eq:FLM_bound}
\end{equation}
where $\sqrt{s_0}$ is an undetermined energy scale, and the coefficient
$B_\text{FLM}$ is also bounded,
\begin{equation}
 B_\text{FLM} \leq \frac{\pi}{m_\pi^2} \approx 60 \text{ mb}\,,
 \label{eq:FLM_bound_B}
\end{equation}
with $m_\pi$ the $\pi^0$ mass. 
The FLM bound and the theory of \textit{Regge poles}
(see, e.g., Refs.~\cite{Collins_book1977,Barone_Predazzi_book2002})
are the inspiring principles behind the currently most successful
phenomenological description of the experimental results for
hadronic total cross sections (see below). 

The recent measurements of $\sigmatot$ at the Large Hadron Collider
(LHC) at CERN have brought back the attention to the energy dependence
of this quantity at high energy, with several new phenomenological
analyses of the data (see Ref.~\cite{Pancheri_Srivastava:2016} and
references therein). Such measurements, made by the TOTEM
Collaboration at $\sqrt{s}=7$ TeV and $\sqrt{s}=8$
TeV~\cite{TOTEM2011b,TOTEM2013a,TOTEM2013c,TOTEM2013d,TOTEM:2016},   
and by the ATLAS Collaboration at $\sqrt{s}=7$ TeV~\cite{ATLAS2014}, 
are currently the highest-energy measurements performed at colliders.
Recently there have also been some advancements on the theoretical side. 
Two of us have obtained the leading behavior of the total
cross section for meson-meson scattering~\cite{Giordano_Meggiolaro2014}
in the framework of the nonperturbative approach to soft high-energy
scattering in QCD proposed long ago by Nachtmann~\cite{Nachtmann1991},
and later developed by several authors~\cite{Dosch_etal1994,Rueter_Dosch1996,Nachtmann1997_1,Berger_Nachtmann1999,Shoshi_etal2002}.
In this approach the relevant scattering amplitudes are related to the
correlation functions (in the sense of the functional integral) of
certain Wilson loops, describing the classical trajectories of the
colliding mesons. It has been argued that the same correlation
functions should be relevant to the description of scattering processes
involving baryons as well~\cite{Rueter_Dosch1996}, so the results of
Ref.~\cite{Giordano_Meggiolaro2014} should apply to general hadronic
processes. Under certain assumptions,  
the leading asymptotic energy dependence turns out to be of the
``Froissart'' type, i.e., proportional to $\ln^2 s$,
with a universal prefactor, independent of the type of particles involved.
More precisely, the first two leading terms in energy read
\begin{equation}
 \sigmatot^{ab}(s)\! \underset{s\to\infty}{\sim}\! B
 \ln^2\left(\frac{s}{s_0^{ab}}\right) + C
 \ln\left(\frac{s}{s_0^{ab}}\right)
 \ln\left[\ln\left(\frac{s}{s_0^{ab}}\right)\right], 
 \label{eq:sigtot_asymp_intro}
\end{equation}
where $\sigmatot^{ab}$ denotes the total cross section for the
scattering of hadrons $a$ and $b$ with masses $m_a$ and $m_b$,
respectively, and $s_0^{ab} = m_am_b$. The coefficients $B$ and $C$
are equal to $\kappa B_\text{th}$ and $\kappa C_\text{th}$, respectively,
where $B_\text{th}$ and $C_\text{th}$ are universal quantities related
to the QCD stable spectrum, while the coefficient $\kappa$ is, at the present
stage, an undetermined real number with $0\le \kappa\le 2$, connected to the
asymptotic ratio between the total elastic cross section ($\sigmael$) and the
total cross section as $\sigmael/\sigmatot=\kappa/2$. As experiments suggest 
universality of the $\ln^2 s$ term, it is natural to \textit{assume} that
$\kappa$ is universal as well. Making more detailed assumptions on
$\kappa$, e.g., that a \textit{black-disk} behavior is attained at high energy
($\kappa=1$), it is then possible to provide a \textit{prediction} for
the coefficient of the leading term in $\sigmatot$, which turns out to
be in fair agreement with the experimental results.
Since a fair amount of assumptions is involved in the
derivation, this result cannot be claimed to be ``the'' prediction of
QCD. Nevertheless, to the best of our knowledge, this is the closest
to a prediction that one has come so far.

We point out that a subleading term of the form $\ln s\cdot \ln\ln s$
has appeared in other approaches to the study of $\sigmatot$, namely
(i) in a bound for the ``energy-averaged'' total cross section
obtained by Martin and Roy~\cite{Martin_Roy2014},
(ii) in a revision of Heisenberg's model made by Nastase and
Sonnenschein~\cite{Nastase_Sonnenschein2015}, where they assume that the
Froissart bound is saturated,
and (iii) in an improvement of the FLM bound in the AdS/CFT approach made
by D\'iez \textit{et al.}~\cite{Diez_etal2015}.

Regardless of its origin, the expression
eq.~\eqref{eq:sigtot_asymp_intro} provides a perfectly good
parametrization of the total cross section at high energy,
that can be used to fit the experimental results, treating $B$ and $C$
as fitting parameters. It is then legitimate to investigate whether in this
way one can improve over the currently most successful parametrizations, where
this term is absent. In particular, given that new experimental information is
now available at higher energies, one may ask whether the asymptotic subleading
contribution in eq.~(\ref{eq:sigtot_asymp_intro}) is already visible at LHC
energies and, if so, how much this contribution is. The main purpose of the
present analysis is precisely to test the functional form
eq.~(\ref{eq:sigtot_asymp_intro}), by means of fits to $pp$ and $\ppbar$ data,
and by comprehensive fits to all the available data for meson-baryon and
baryon-baryon scattering.

Besides this general purpose, our analysis allows to test the viability of the
various approaches leading to an expression like
eq.~\eqref{eq:sigtot_asymp_intro} for the total cross section.
In particular, as already mentioned above, in the approach of
Ref.~\cite{Giordano_Meggiolaro2014} $B$ and $C$
are related to the QCD spectrum and to the
elastic-to-total cross section ratio: this puts severe constraints
on the allowed values of $B$ and $C$.
In pursuing this line of investigation, the coefficients $B$ and
$C$ can either be fixed to their theoretical prediction, or treated as
free parameters when fitting the experimental data.

The paper is organized as follows. In section \ref{sec:theory} we
present a summary of some useful basic relations, and the results of
Ref.~\cite{Giordano_Meggiolaro2014} that constitute the basis of the 
present analysis.
In section \ref{sec:parametrization} we discuss the
parametrization used in the fits and in section
\ref{sec:dataset_methodology} we detail the dataset considered. In
section \ref{sec:results} we present and discuss the results obtained
in the fits. Finally, we present our conclusions in section
\ref{sec:conclusions}.


\section{Summary of Theoretical Results}
\label{sec:theory}

In this section we present a brief summary of the theoretical results of
Ref.~\cite{Giordano_Meggiolaro2014} concerning the asymptotic high-energy
behavior of the elastic hadron-hadron scattering amplitude.

The approach employed in Ref.~\cite{Giordano_Meggiolaro2014} is that
originally proposed by Nachtmann~\cite{Nachtmann1991} for elastic
quark-quark scattering, and later expanded by several authors to
describe hadron-hadron scattering~\cite{Dosch_etal1994,Rueter_Dosch1996,
Nachtmann1997_1,Berger_Nachtmann1999,Shoshi_etal2002}. In this
approach, the elastic scattering amplitude in the soft high-energy
regime ($s\to\infty$, $|t|\lesssim 1~{\rm GeV}^2$) of two mesons $a$
and $b$ with masses $m_a$ and $m_b$, respectively, is obtained from
the elastic scattering amplitude of two colorless dipoles of fixed
transverse size, after folding with appropriate wave functions
describing the colliding mesons. In turn, the dipole-dipole scattering
amplitude is obtained from the (normalized, connected)
correlation function of two Wilson loops in Minkowski space, running 
along the classical trajectories of the dipoles, in the limit of
infinite longitudinal extension of the loops. Exploiting analytic
continuation~\cite{Meggiolaro1997,Meggiolaro1998,Meggiolaro2002,Meggiolaro2005,Giordano_Meggiolaro:2006,Meggiolaro:2007,Giordano_Meggiolaro2009},
it is possible to reconstruct
these correlation functions from their Euclidean counterparts, which
in turn can be related to the QCD spectrum by appropriately inserting
a complete set of states between the two Wilson-loop operators. 
Under certain analyticity assumptions, discussed in detail in
Ref.~\cite{Giordano_Meggiolaro2014}, it is then possible to derive the
asymptotic behavior of the Minkowskian correlators at large energy and
large impact parameter (which is the transverse distance between the
classical trajectories of the incident particles), from which the elastic
amplitude and the total 
cross section are finally obtained. The same chain of arguments can be
used in the case of processes involving baryons, since similar
Wilson-loop correlation functions are involved in the calculation of
the corresponding scattering amplitudes~\cite{Rueter_Dosch1996}, and
the detailed transverse geometry of the Wilson loops plays no role in
the derivation of Ref.~\cite{Giordano_Meggiolaro2014}. The results
discussed below are therefore expected to be valid in more general
hadronic processes than just meson-meson scattering. For more details,
we invite the interested reader to confer the original references.

In Ref.~\cite{Giordano_Meggiolaro2014} the elastic scattering
amplitude was computed assuming that for asymptotically large energies
one finds a \textit{black-disk} behavior, corresponding to the Minkowskian
Wilson-loop correlator tending to zero as the energy goes to infinity,
for sufficiently large but fixed impact parameter. It is
straightforward to generalize this result to the case where this
correlator tends to a nonzero constant $\kappa-1$, which we
\textit{assume} to be independent of the transverse size of the
loops. Analiticity requires $\kappa$ to be real, and unitarity then 
requires that $\kappa\in[0,2]$. One then finds
\begin{equation}
 F^{ab} (s,t) \underset{s\to\infty,\,t\to 0}{\sim} 4\pi i s \kappa
 \left(\frac{\eta}{\tilde{m}}\right)^2 \frac{J_1(x)}{x}\,, 
 \label{eq:elastic_amplitude_GM2014}
\end{equation}
if $x \equiv \eta\sqrt{-t}/\tilde{m}$ is kept fixed. Here 
\begin{equation}
 \eta = \frac{1}{2}W(2e^{2(\tilde{s}-1)\chi}) = (\tilde{s}-1)\chi
 -\frac{1}{2} \ln[(\tilde{s}-1)\chi] + 
 \frac{\ln[(\tilde{s}-1)\chi]}{4(\tilde{s}-1)\chi} + \dots \,,
 \label{eq:def_eta}
\end{equation}
with $W$ the Lambert $W$ function~\cite{LambertW}, and $\chi =
\ln(s/s_0^{ab})$ with $s_0^{ab}=m_am_b$. In the
equations above, $\tilde{s}$ and 
$\tilde{m}$ are, respectively, the spin and mass of the particle in
the QCD stable spectrum that maximizes the ratio  
\begin{equation}
 l_p \equiv \frac{s_p-1}{m_p} \quad (s_p>1)\,,
\label{eq:ratio_s_m}
\end{equation}
where $s_p$ and $m_p$ are the spin and the mass of particle $p$.
By ``QCD stable spectrum'' we mean here all those particles that are
stable when strong interactions are considered in isolation. 
The assumed independence of $\kappa$ on the transverse size of the dipoles
implies that the scattering amplitude
eq.~\eqref{eq:elastic_amplitude_GM2014}, obtained after trivially folding
with the hadronic wave functions, is a universal function of $x$.

Taking the limit $t\to0$ in
eq.~\eqref{eq:elastic_amplitude_GM2014} and using the optical theorem
eq.~\eqref{eq:optical_theorem}, we get for the total cross section 
\begin{equation}
 \sigmatot^{ab}(s) \to \frac{2\pi}{\tilde{m}^2}\kappa\eta^2  +
 \mathcal{O}(\eta)\,. 
 \label{eq:sigmatot_case2}
\end{equation}
Using now eq.~\eqref{eq:def_eta}, we obtain, up to first subleading order,
\begin{equation}
 \sigmatot^{ab}(s) \to \kappa  \left\{
   B_\text{th}\ln^2\left(\frac{s}{s_0^{ab}}\right)
 + C_\text{th} \ln\left(\frac{s}{s_0^{ab}}\right)
   \ln\left[\ln\left(\frac{s}{s_0^{ab}}\right) \right]\right\}
 + \mathcal{O}\left(\ln\left({s}/{s_0^{ab}}\right)\right)\,,
 \label{eq:sigmatot_th} 
\end{equation}
with
\begin{equation}
 B_\text{th} = 2\pi\frac{(\tilde{s}-1)^2}{\tilde{m}^2}\,,~~~~
 C_\text{th} = - 2\pi\frac{(\tilde{s}-1)}{\tilde{m}^2}\,.\label{eq:B_C_th}
\end{equation}
Notice the relations
\begin{equation}
 \frac{B_\text{th}}{C_\text{th}} = 1-\tilde{s}\,,\qquad
2\pi\frac{B_\text{th}}{C_\text{th}^2} = \tilde{m}^2\,.
 \label{eq:ratio_BC}
\end{equation}
We then find that in the limit $s\to\infty$ both the leading and the subleading
term in the total cross section are independent of the scattering particles.

We also want to mention that in Ref.~\cite{Giordano_Meggiolaro2014} 
another subleading term of order $\mathcal{O}(\ln s)$, i.e., $Q^{ab}\ln
(s/s_0^{ab})$, was also found (and such a term is also present in the analysis
of Ref. \cite{Diez_etal2015}). Contrary to the leading and subleading
terms discussed above, the coefficient $Q^{ab}$ is expected to depend
on the colliding particles, even if our assumption on $\kappa$ is met.

In Ref.~\cite{Giordano_Meggiolaro2014} the values of $B_\text{th}$ and
$C_\text{th}$ were estimated by maximizing the ratio
eq.~\eqref{eq:ratio_s_m} over the higher-spin, QCD stable
spectrum. The resulting ``dominant'' particle was found to be
the $\Omega^\pm$ baryon, with mass $m_{\Omega^{\pm}} \approx 1.67$~GeV
and spin $3/2$, that yields [using eq.~\eqref{eq:B_C_th}] 
\begin{equation}
 B_\text{th}^{\Omega}  = 0.22 \text{ mb},~~~~
 C_\text{th}^{\Omega}  = -2B_\text{th}^\Omega = -0.44 \text{ mb}.
 \label{eq:B_C_th_Omega}
\end{equation}
One of the assumptions in Nachtmann's approach is that the processes
of splitting and annihilation of partons inside hadrons can be
neglected over a small time window around collision
time~\cite{Nachtmann1991}. 
The description of hadrons in terms of dipoles is therefore perhaps
better justified in the \textit{quenched} limit of the theory. In this
case the relevant spectrum over which one has to maximize
eq.~\eqref{eq:ratio_s_m} is the \textit{glueball} spectrum of the
pure-gauge theory. The states of interest here (for details see
Ref.~\cite{Giordano_Meggiolaro2014}) are the $J^{PC}=2^{++}$ glueball state,
with mass $m_{g2^{++}} \approx 2.40$ GeV, and the $3^{+-}$, with mass
$m_{g3^{+-}} \approx 3.55$ GeV, both calculated in the
\textit{quenched} approximation~\cite{Morningstar_Peardon1999}, for
which one finds
\begin{equation}
 B_\text{th}^{g2^{++}} = 0.42 \text{ mb}\,,~~~~
 C_\text{th}^{g2^{++}} = -B_\text{th}^{g2^{++}} = -0.42 \text{ mb}\,,
 \label{eq:B_C_th_g2pp}
\end{equation}
and
\begin{equation}
 B_\text{th}^{g3^{+-}} = 0.78 \text{ mb}\,,~~~~
 C_\text{th}^{g3^{+-}} = -\frac{1}{2}B_\text{th}^{g3^{+-}} = -0.39 \text{ mb}\,.
\label{eq:B_C_th_g3pm}
\end{equation}
The value of $B_\text{th}$ obtained using the $\Omega^{\pm}$ baryon is
the closest to the values of $B$ published in the Review of Particle
Physics by the Particle Data Group (PDG).
In the 2014 edition, fits to forward quantities using energies $\sqrt{s}\geq 5$
GeV yielded $B_\text{PDG}~=~0.2704~\pm~0.0038$~mb~\cite{PDG2014};
in the 2016 edition the revised value is $0.2720~\pm~0.0024$~mb,
obtained from a fit to data with $\sqrt{s} \geq 5$ GeV
and including the most recent measurements at the LHC~\cite{PDG2016}. 

From the amplitude eq.~\eqref{eq:elastic_amplitude_GM2014}
one can easily calculate the asymptotic behavior of the total elastic cross 
section $\sigmael^{ab}$ using
\begin{equation}
 \sigmael^{ab}(s) = \int_{-\infty}^0 dt \frac{d\sigmael^{ab}}{dt}(s,t),~~~~
 \frac{d\sigmael^{ab}}{dt}(s,t) = \frac{1}{16\pi s^2}|F^{ab}(s,t)|^2,
 \label{eq:sigmael_def}
\end{equation}
and assuming that the small-$t$ region gives the dominant contribution.
From this result and from eq.~\eqref{eq:sigmatot_th}, we obtain 
\begin{equation}
 \frac{\sigmael^{ab}}{\sigmatot^{ab}} \sim \frac{\kappa}{2} \quad (s\to\infty).
 \label{eq:ratio_eltot_kappa}
\end{equation}
Eqs. \eqref{eq:elastic_amplitude_GM2014} and \eqref{eq:ratio_eltot_kappa}
show that, if $\kappa<1$ (i.e., $\sigmael/\sigmatot < 1/2$),
the elastic scattering amplitude $F^{ab}(s,t)$
behaves asymptotically as a \textit{grey disk} with constant profile function
equal to $\kappa$ and radius growing as $\ln s$.\footnote{We remind
  the reader that if one expresses the elastic amplitude $F(s,t)$ in
  terms of the \textit{profile function} $\Gamma(s,b)$, i.e. (assuming
  azimuthal symmetry)~\cite{Barone_Predazzi_book2002,Block_Cahn1985},
  $F(s,t) =4\pi i s\int_0^\infty  db\,b\, J_0(b\sqrt{-t})\Gamma(s,b)$,
  with $b$ the absolute value of the impact parameter,
  then the so-called \textit{grey disk} of radius $R(s)$ is a simple model
  in which $\Gamma_\text{GD}(s,b) = \Gamma_0(s)$ for $0 \leq b \leq R(s)$,
  and $\Gamma_\text{GD}(s,b) = 0$ for $b > R(s)$,
  which leads to $F_\text{GD}(s,t) =
  4\pi i sR(s)\Gamma_0(s) {J_1(R(s)\sqrt{-t})}/{\sqrt{-t}}$
  and, using eq.~\eqref{eq:sigmael_def} and the optical theorem
  eq.~\eqref{eq:optical_theorem}, to $\sigmael/\sigmatot =
  \Gamma_0(s)/2$.} 
If $\kappa=1$ we have the so-called \textit{black disk}
($\sigmael/\sigmatot = 1/2$),
and if $\kappa > 1$ we have the \textit{antishadowing} regime
($\sigmael/\sigmatot >1/2$)~\cite{Troshin_Tyurin1993,Troshin_Tyurin2007}.  

We want to stress the fact that so far we have assumed that $\kappa$
is a universal parameter.
Although this assumption is made plausible by the
observed universality of total cross sections at high energy, 
a more general analysis should start with a particle-dependent parameter,
$\kappa^{ab}$, and universality should then emerge from the fit results
rather than being imposed \textit{a priori}. This would however
increase considerably the number of fitting parameters.
Therefore, even if this case is not the most general
one, we will consider $\kappa$ as a universal parameter in the
present analysis.


\section{Parametrization of the Total Cross Section}
\label{sec:parametrization}

In this work we want to study the leading and subleading high-energy
behavior of hadron-hadron total cross sections using a best-fit
analysis of hadronic scattering data.
The available data of highest energy (excluding cosmic rays) are those
for $pp$ scattering at 8 TeV at the LHC~\cite{TOTEM2015a,TOTEM:2016,ATLAS2016}
and for $\bar{p}p$ scattering at 1.8 TeV at the Tevatron.
If we want to include also data from other 
meson-baryon and baryon-baryon scattering processes
(which are available only for $\sqrt{s} \lesssim 25$~GeV),
we need a parametrization for the low-energy region as well.
In this section we present our parametrization for the total cross section
and we fix the notation for the best-fit variants considered in our analysis. 

We will divide the total cross section in two parts, 
aimed at describing the low-energy (LE) and the high-energy (HE)
regions, respectively:
\begin{equation}
 \sigmatot(s) = \sigma_\text{LE}(s)+ \sigma_\text{HE}(s).
 \label{eq:sigtot_par_notation}
\end{equation}
As usual in the literature, we parametrize the energy dependence of
$\sigmatot(s)$ at low energy by means of \textit{Reggeon} exchange
in the $t$-channel \cite{Collins_book1977,Barone_Predazzi_book2002}.
The Reggeon-exchange contribution reads
\begin{equation}
 R_i^{ab}(s) = A_i^{ab}\left(\frac{s}{s^{ab}_0}\right)^{-b_i} \quad(i=1,\,2)\,,
\end{equation}
where $A_i^{ab}$ is associated to the residue function and $b_i$
to the intercept of the Reggeon trajectory $\alpha_i(t)$, i.e., $b_i = 1-
\alpha_i(0)$. We consider two Reggeon contributions: the index $i=1$
corresponds to a trajectory with \textit{even} signature, while $i=2$
corresponds to a trajectory with \textit{odd} signature.
The latter contributes with a \textit{minus} sign to $ab$ scattering and with
a \textit{plus} sign to the crossed channel, $\bar{a}b$. Summarizing,
the low-energy parametrization reads\footnote{In the left-hand side of
the equation, $a^+ \equiv a$ represents a positive-charge particle, while
$a^-\equiv \bar{a}$ corresponds to its negative-charge antiparticle.} 
\begin{equation}
 \sigma_\text{LE}^{a^\pm b}(s) =
 A_1^{ab}\left(\frac{s}{s_0^{ab}}\right)^{-b_1} \mp
 A_2^{ab}\left(\frac{s}{s_0^{ab}}\right)^{-b_2} \,.
 \label{eq:sigtot_le}
\end{equation}
The high-energy contribution is parametrized by adding to
the leading ($\sim \ln^2 s$) and subleading ($\sim \ln s \cdot \ln\ln
s$) terms also a constant term $\AP^{ab}$ (which is usually understood
as the contribution of the \textit{critical Pomeron}, i.e., a simple
pole in the complex angular-momentum plane with intercept
$\alpha_\mathbb{P}(0)=1$);\footnote{
\label{foot:const_term} 
Actually, a constant term $\AP^{ab}$ could also originate from the sum of the
\textit{critical} (i.e., \textit{simple-pole}) Pomeron with the constant term
generated by higher-order singularities, e.g., a \textit{triple-pole} Pomeron
(which also gives a $\ln^2 s$ term). In any case, however, its value is
obviously affected by the choice of the energy scale in the leading and
subleading terms in eq. \eqref{eq:sigtot_he} (see also the discussion at the
end of this section).}
i.e.:
 \begin{equation}
 \sigma_\text{HE}^{a^\pm b}(s) = \AP^{ab} + \kappa\left\{B\ln^2\left(\frac{s}{s_0^{ab}}\right)
 + C\ln\left(\frac{s}{s_0^{ab}}\right)\ln\left[\ln\left(\frac{s}{s_0^{ab}}\right)\right]\right\}\,, 
 \label{eq:sigtot_he}
 \end{equation}
where, following the analysis of Ref.~\cite{Giordano_Meggiolaro2014}
(briefly recalled in sec. \ref{sec:theory}), $B$ and $C$ are treated as
\textit{universal} parameters, 
while $\AP^{ab}$, as well as $A_1^{ab}$ and $A_2^{ab}$,
are \textit{reaction-dependent}.
As already said at the end of sec. \ref{sec:theory}, we shall also assume
(as in Ref.~\cite{Giordano_Meggiolaro2014}) that $\kappa$
is independent of the properties of the scattering particles.
We consider the energy scale to be a fixed parameter and to depend
only on the masses of the scattering particles, $s_0^{ab}=m_am_b$.

Summarizing, $A_i^{ab}$ (mb), $b_i$ (dimensionless) and $\AP^{ab}$
(mb) are always free parameters to be determined in the fits. The
parameters $B$ (mb), $C$ (mb) and $\kappa$ (dimensionless) can be
fixed or free, as detailed below in the descriptions of our variants
of fits. 

The names of the variants are written using the following notation: LT
stands for \textbf{L}eading \textbf{T}erm, SLT for \textbf{S}ub
\textbf{L}eading \textbf{T}erm, the subscript ``th'' refers to the
case where we fix $B$ ($\LTth$) or both $B$ and $C$ ($\SLTth$)
to the theoretical values discussed in section \ref{sec:theory},
and the subscript ``$\kappa$'' indicates that the parameter $\kappa$ is free.
The variants considered here are the following:

 \begin{itemize}
  \item $\LT$: $C=0$ and $\kappa = 1$ are fixed parameters, while $B$ is
    free. This case (in which the subleading term is absent) corresponds
    essentially to the parametrization used by the PDG in their analysis of
    forward data~\cite{PDG2016} and to the highest-rank result
    obtained by the COMPETE Collaboration~\cite{COMPETE2002} (see
    below for more details).
  
  \item $\SLT$: $\kappa = 1$ is fixed, while $B$ and $C$ are free
    parameters. {This case corresponds to the previous parametrization
      with the inclusion of the subleading term.}
  
  \item $\LTth$: $B=B_\text{th}$, $C=0$ and $\kappa=1$ are fixed parameters.
    This variant has $B$ fixed to the theoretical values discussed in
    sec.~\ref{sec:theory} and no subleading term.
  
  \item $\SLTth$: $B=B_\text{th}$, $C=C_\text{th}$ and $\kappa=1$
    are fixed parameters.
    This variant has both $B$ and $C$ fixed to the theoretical values
    discussed in sec.~\ref{sec:theory}.

  \item $\SLTthk$: $B=B_\text{th}$ and $C=C_\text{th}$ are fixed
    parameters, while $\kappa$ is a free parameter.
 
 \end{itemize}

 The main difference between our $\LT$ parametrization, the
 highest-rank result by COMPETE, and the PDG parametrization, is in the
 energy scale appearing in the leading term $\ln^2 s$.
 In the COMPETE analysis, the energy scale $s_0$ is a free parameter,
 which does not depend on the scattering particles.
 Our energy scale, on the other hand, is fixed and depends only on the
 masses of the scattering particles, $s_0^{ab} =  m_am_b$.
 In the PDG analysis this scale depends on the masses of
 the colliding particles and on a universal mass scale also entering
 their parametrization of the coefficient $B$, so it contains both a
 fixed and a free part. All in all, this results in the three
 parametrizations differing pairwise by a particle-dependent
 term of order ${\cal O}(\ln s)$ (besides a particle-dependent
 constant term that can be reabsorbed in $\AP^{ab}$). Since very
 high-energy data are not yet available, this 
 can affect the result obtained for $B$ in the fits. 


\section{Dataset and Methodology}
\label{sec:dataset_methodology}

In this section we describe our dataset and the methodology used in our fits.

\subsection{Dataset}
\label{sec:dataset}

Our dataset comprises data from meson-baryon and baryon-baryon
scattering, namely $\pp$, $\ppbar$, $pn$, $\bar{p}n$, $\pi^{\pm}p$,
$K^{\pm}p$, $K^{\pm}n$, with cutoff energy
$\sqrt{s_\text{min}} = 5$ GeV (so that we are well above the resonance region).
This cutoff energy is the same used in the highest-rank result by 
the COMPETE Collaboration~\cite{COMPETE2002} and in the PDG
analysis~\cite{PDG2016}. Only data obtained in accelerator experiments
were included in the fits, i.e., no cosmic rays data were considered,
and the datasets are those available at the PDG
website~\cite{PDG_data_website}.  

For $\pp$ scattering, besides the data already in the PDG dataset at 7 TeV,
we have included further data at 7 and 8 TeV obtained by the TOTEM and ATLAS
Collaborations, namely,
the luminosity-independent measurement at 7 TeV~\cite{TOTEM2013c},
the first measurement at 8 TeV~\cite{TOTEM2013d} and the values of
$\sigmatot$ obtained together with the $\rho$ value at 8 TeV from the
Coulomb-Nuclear interference region in the differential cross section
data~\cite{TOTEM:2016} by the TOTEM Collaboration, and the measurement
at 7 TeV by the ATLAS Collaboration~\cite{ATLAS2014}. 
This information is summarized in table \ref{tab:dataLHC}.
Our dataset therefore comprises data with energy
in the range 5~GeV~$\leq~\sqrt{s}~\leq$~8~TeV. 

Cosmic-ray data~\cite{ARGO_YBJ2009,PierreAuger2012,TA:2015} 
are shown in the figures just to illustrate the
trend with energy and, as stated above, they have not been included in
the fits. Since their uncertainties are large, we expect that their inclusion
in the fits would not change much the results.
The energy range and the number of points available for each scattering
channel that we have considered are shown in table \ref{tab:info_datasets}. 
 

\begin{table}
\caption{\label{tab:dataLHC}Experimental data of $\sigmatot$ in the
  LHC energy region.}    
\begin{tabular}{c c c c c}\hline\hline
$\sqrt{s}$ (TeV) & $\sigmatot$ (mb) & Collaboration \\
\hline
7 & 98.3  $\pm$ 2.8 & TOTEM \cite{TOTEM2011b} \\
7 & 98.6  $\pm$ 2.2 & TOTEM \cite{TOTEM2013a} \\
7 & 98.0  $\pm$ 2.5 & TOTEM \cite{TOTEM2013c} \\
7 & 99.1  $\pm$ 4.1 & TOTEM \cite{TOTEM2013c} \\
7 & 95.4  $\pm$ 1.4 & ATLAS \cite{ATLAS2014}  \\\hline
8 & 101.7 $\pm$ 2.9 & TOTEM \cite{TOTEM2013d} \\
8 & 102.9 $\pm$ 2.3 & TOTEM \cite{TOTEM:2016} \\
8 & 103.0 $\pm$ 2.3 & TOTEM \cite{TOTEM:2016} \\\hline\hline
\end{tabular}
\end{table}

 In all the cases, we have treated the datapoints as independent, 
 including those that have the same energy. For all data we have considered
 statistic and systematic uncertainties added in quadrature. 

 We stress that we are not including data from reactions that involve photons
 or deuterons and we do not constrain our fits using the data for the $\rho$
 parameter, as it is done by COMPETE and PDG.
 
 Finally, we mention that there are 9 points available 
 for $\Sigma^-p$ scattering in the energy region of
 interest~\cite{PDG_data_website}.   
 Including these points makes the fits more unstable (due to the
 absence of data in the corresponding crossed channel), but
 this does not affect the results, because of their large errors.
 We have therefore decided not to include them in our analysis. 

\begin{table}
\caption{\label{tab:info_datasets} Information about the reactions
  in our dataset: minimum energy, maximum energy and
  number of points for each reaction.} 
\begin{tabular}{c c c c}\hline\hline
 Reaction    & $\sqrt{s_\text{min}}$ (GeV) & $\sqrt{s_\text{max}}$ (GeV) & \# points\\\hline
 $\pp$        & 5.01 & 8000  & 112\\
 $\ppbar$  & 5.16 & 1800  & 59\\
 $pn$        & 5.30 & 26.40 & 34\\
 $\bar{p}n$  & 5.18 & 22.98 & 33\\\hline
 $\pi^+p$    & 5.21 & 25.28 & 50\\
 $\pi^-p$    & 5.03 & 34.67 & 95\\
 $K^+p$      & 5.13 & 24.14 & 40\\
 $K^-p$      & 5.11 & 24.14 & 63\\
 $K^+n$      & 5.24 & 24.16 & 28\\
 $K^-n$      & 5.11 & 24.16 & 36\\\hline
             &      & Total:& 559\\\hline\hline
\end{tabular}
\end{table} 
 
\subsection{Methodology}
\label{sec:methodology}

We first consider fits using 
our parametrization, eqs.~\eqref{eq:sigtot_par_notation} and
\eqref{eq:sigtot_le}--\eqref{eq:sigtot_he},
to data from $pp$ and $\ppbar$ scattering only. This pair of reactions
constitutes the set with available data in the largest energy
range. Since the subleading term of interest here is expected to be
relevant at high energies, it is important to estimate its
contribution without much weight from the low-energy data from other
reactions. Nevertheless, we have also considered fits to all the data
for the hadron-hadron processes shown in table \ref{tab:info_datasets}.

\begin{figure}[h!]
\centering
\includegraphics[scale=1]{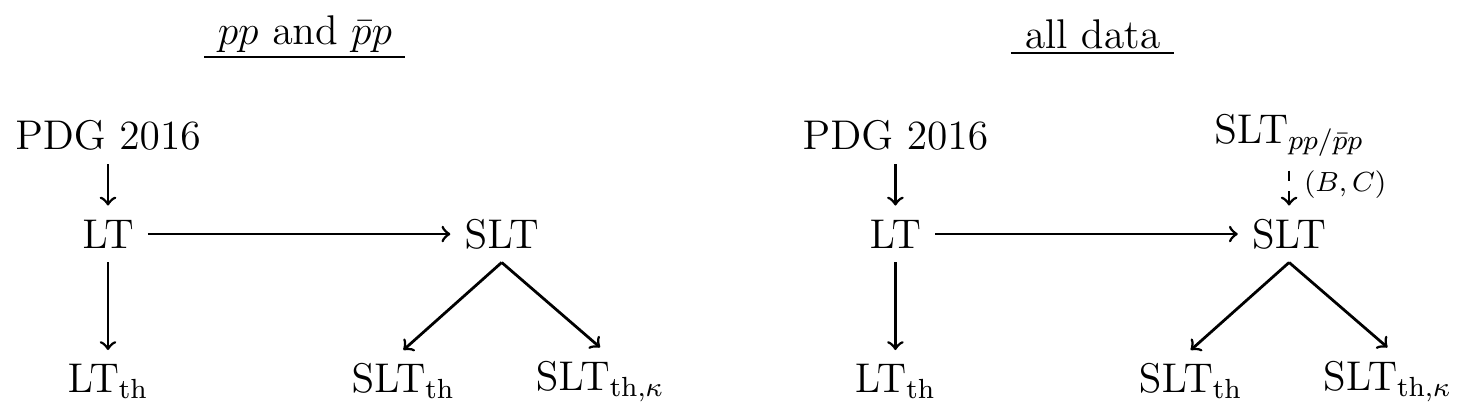}
\caption{\label{fig:IV_scheme} Initial-value scheme used in the fits.}
\end{figure}

In order to start from a solid and updated result, we decided to use
as initial values for the $\LT$ fit the results presented in PDG
2016~\cite{PDG2016}, and 
then use the results of $\LT$ as initial values for the $\SLT$ fit. In
this way, fitting first $\LT$ (that essentially corresponds to the PDG
parametrization, as we have already observed above) we create a
reference for discussing differences when we include the subleading
term as well ($\SLT$), instead of comparing directly to the PDG
result. This procedure, however, turned out to be problematic in the 
fit to all hadronic data (see below
sec.~\ref{sec:res_allreactions_SLT}). In that case we therefore
decided to use as initial values for the parameters $B$ and $C$ in
the $\SLT$ fit to all data the results obtained in the $\SLT$ fit to
$\pp$/$\ppbar$ data. For the other parameters, we used the results
obtained in the $\LT$ fit to all data. The detailed 
scheme is shown in fig.~\ref{fig:IV_scheme}, where $X \rightarrow
Y$ means that the results of variant $X$ were used as
initial values for the fit with variant $Y$.

We use the reduced chi-squared ($\chi^2/\nu$, where $\nu$ is the
number of degrees of freedom) as a measure of the quality of the 
fit~\cite{MINUIT_Manual1994,Bevington_Robinson_book1992}.
However, we do not base our preference for a result over another
on small differences in the value of this quantity,
since the inclusion of both statistical and systematic
uncertainties puts some limits on the use of this test.
It is important to note that some data obtained by the TOTEM Collaboration
do not have statistical uncertainty (as can be seen in the caption of
table 1 in Ref.~\cite{TOTEM2013c}).

The fits were performed with the class TMinuit from the ROOT
Framework~\cite{ROOT_website} with $1\sigma$ of confidence level. 


\section{Fit Results}
\label{sec:results}

In this section we present the results obtained in our fits, first to
$\pp$ and $\ppbar$ data only, and later to all reactions, considering
all variants described above in sec.~\ref{sec:parametrization}.
Finally, we compare and discuss our results.

\subsection{Fits to $\pp$ and $\ppbar$ Data}
\label{sec:results_fits_pp_ppbar}

\begin{figure}[h!]
\centering
\includegraphics[scale=0.45]{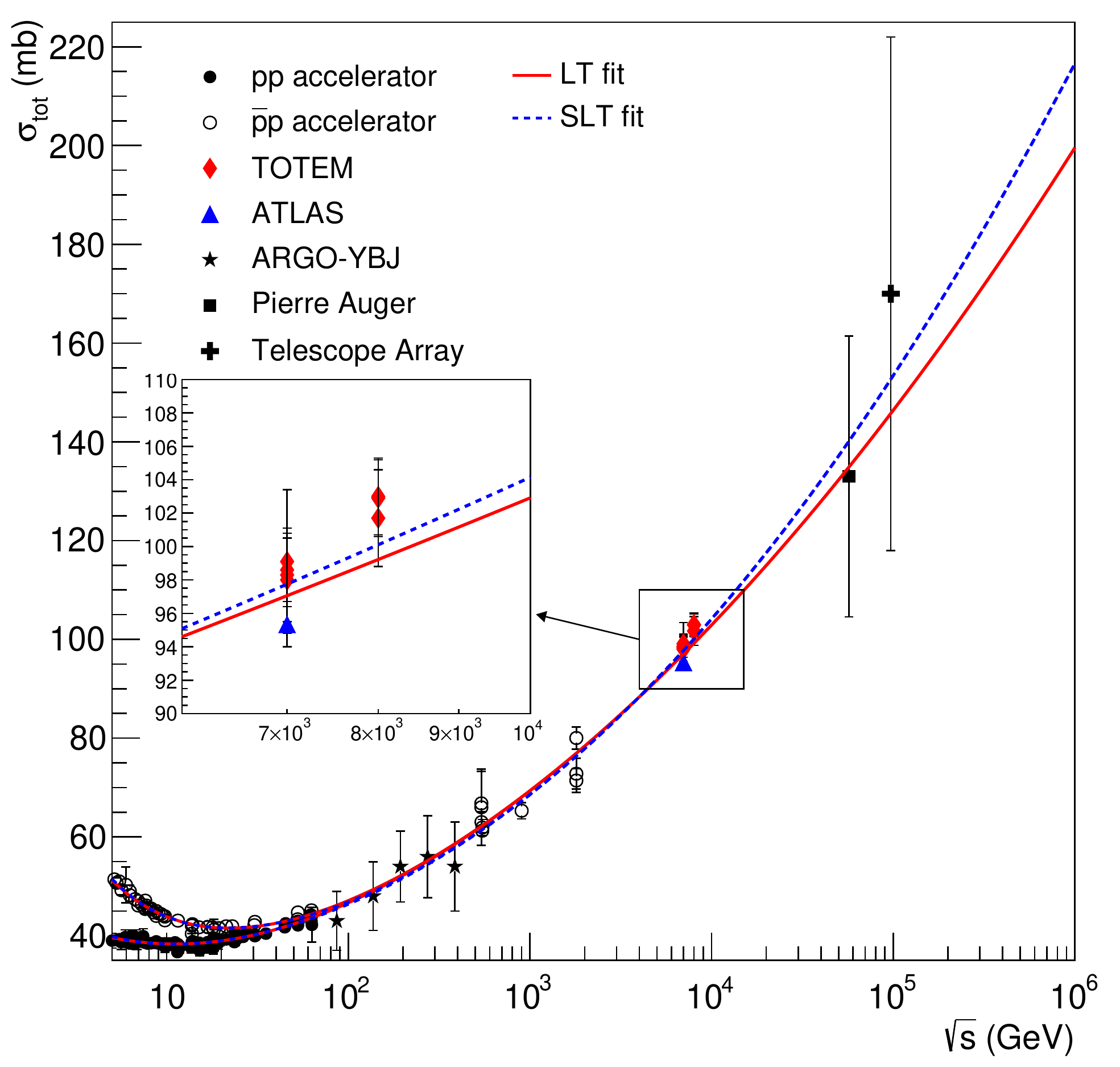}
\caption{\label{fig:res_pp_pbarp}Results of fits with $\LT$ and
 $\SLT$ to $pp$ and $\ppbar$ data.} 
\end{figure}

The parameters obtained in fits to $\pp$ and $\ppbar$ data with $\LT$ and
$\SLT$ are shown in table \ref{tab:res_pp_pbarp}, those with $\LTth$,
$\SLTth$ and $\SLTthk$ in table \ref{tab:res_pp_pbarp_SLTkf}.
The curves calculated with these parameters are compared with the
experimental data in figs.~\ref{fig:res_pp_pbarp},
\ref{fig:res_pp_pbarp_SLTf} and \ref{fig:res_pp_pbarp_SLTkf}.
Below we discuss the results that we have obtained using the different
variants.

\subsubsection{Fits with $\LT$ and $\SLT$}
\label{sec:res_pp_pbarp_SLT}

The results obtained with variants $\LT$ and $\SLT$
provide a good description of the experimental data.
Although there is a small decrease in 
the value of $\chi^2/\nu$ going from $\LT$ to $\SLT$, nevertheless, as
discussed in section \ref{sec:methodology}, we cannot favor one
variant over the other on the basis of this value. Given
that both are $\lesssim 1$, we can say that both variants 
result in good fits to the data. 

Interestingly enough, the $\LT$ fit yields for $B$ a value close to
the theoretical prediction $B_\text{th}$ obtained in
Ref.~\cite{Giordano_Meggiolaro2014} using the mass and spin of the
$\Omega^\pm$ baryon.

With $\SLT$ we have obtained $C<0$ and $C\neq0$ within the uncertainty.
The negative value of $C$ causes an increase in the $B$
parameter and also in $\AP$ to compensate for the negative
contribution of the subleading term. 
The uncertainty on $\AP$ increases one order of magnitude, but the relative
uncertainty is still small ($\sim$ 5.6\%). 

Given the small difference mentioned above between the values of
$\chi^2/\nu$ obtained with $\LT$ and $\SLT$, we cannot claim that the fit
with the subleading term represents an improvement over the fit without it.
We can only say that the data are compatible with a nonzero value of $C$.

Regarding the Reggeon trajectories, the values of $b_1$ and $b_2$ obtained
in $\LT$ are not far from the values obtained by the PDG and also in other
analyses (such as, for example, the one by Menon and Silva
in Ref.~\cite{Menon_Silva2013b}).
When changing from $\LT$ to $\SLT$, i.e., allowing $C$ to be nonzero,
we find that $A_2$ and $b_2$ are practically stable, while $A_1$ and $b_1$
increase; moreover, we observe that the values of $b_1$ and $b_2$ are
compatible within the errors, i.e., the intercepts of the two Reggeon
trajectories become degenerate.
A similar effect was observed by COMPETE in Ref.~\cite{COMPETE2002},
when discussing their highest-rank result (similar to $\LT$).
In that case, the degeneracy of the Reggeon intercepts was ascribed to
a decreasing contribution of the log-squared term for $s<s_h$, where
$s_h$ is the energy scale determined in the fit. 

\begin{table}
  \caption{\label{tab:res_pp_pbarp}Results of fits with $\LT$ and
    $\SLT$ to $\sigmatot$ data of $\pp$ and $\ppbar$ scattering.
    Parameters $A_1$, $A_2$, $\AP$, $B$, $C$
    are in mb, while $b_1$, $b_2$ and $\kappa$ are dimensionless.} 
  \begin{tabular}{c|cc}\hline\hline
               & \multicolumn{2}{c}{Fits to $\sigmatot$}  \\\hline
               & $\LT$  & $\SLT$   \\\hline
  $B$          & 0.2269(38) &  0.349(29)   \\ 
  $C$          & 0 (fixed)  & -0.95(21)    \\
  $b_1$        & 0.342(15)  &  0.560(76)   \\
  $b_2$        & 0.539(15)  &  0.541(16)   \\
  $\kappa$     & 1 (fixed)  &  1 (fixed)   \\
  $A_1$        & 56.8(1.7)  &  64.4(8.2)   \\
  $A_2$        & 35.2(2.5)  &  35.6(2.5)   \\
  $\AP$        & 24.77(60)  &  35.7(2.0)   \\\hline
  $\chi^2/\nu$ & 0.972      &  0.933       \\
  $\nu$        & 165        &  164         \\\hline\hline
  \end{tabular}                                          
\end{table}

In Ref.~\cite{Diez_etal2015}, D\'iez, Godbole and Sinha performed fits
to $\sigmatot$ data from $pp$ and $\ppbar$ scattering in order to
determine what is the dominant subleading contribution to $\sigmatot$
between $\ln s$ and $\ln s \ln\ln s$ (see section \ref{sec:theory}).
They also parametrize the low-energy
contributions to $\sigmatot$ with Reggeon terms, however 
fixing some of the parameters according to the expectations from Regge
theory, and some others using sum rules in the resonance region
(following the works by Block and Halzen~\cite{Block_Halzen2006}).
Their analysis favors the $\ln s$ term over the $\ln s\ln\ln s$ one.
Their approach differs from ours mainly in two aspects: (i) they fix
low-energy parameters, which we do not, and (ii) they do not address
the question of the dependence on the colliding particles, while in
Ref.~\cite{Giordano_Meggiolaro2014} it is argued that the $\ln s
\ln\ln s$ is universal, while the $\ln s$ term is reaction dependent.

In the present analysis we want to investigate only the universal
terms, leaving the $\ln s$ contribution to a future study. In
fact, the inclusion of this kind of term would increase the number of
free parameters, especially in the fits to all reactions.
We think that for the time being it is better to keep the fits as simple
as possible and avoid a large number of parameters.

\subsubsection{Fits with $\LTth$ ($\kappa=1$, $B=B_\text{th}$, $C=0$),
$\SLTth$ ($\kappa=1$, $B=B_\text{th}$, $C=C_\text{th}$) and
$\SLTthk$ ($\kappa$ free, $B=B_\text{th}$, $C=C_\text{th}$)}
\label{sec:res_pp_pbarp_SLTf}

After performing the fits with $\LT$ and $\SLT$, we did fits where $B$
and $C$ were fixed {to the values} discussed in section
\ref{sec:theory}. In table \ref{tab:res_pp_pbarp_SLTkf} we
divide the results according to the variants and to the values of $B$
and $C$ used.

\paragraph{Fits with $\LTth$ and $\SLTth$: $B$ and $C$ from the $\Omega^\pm$ baryon.}
  
\begin{figure}[h!]
\centering
\includegraphics[scale=0.45]{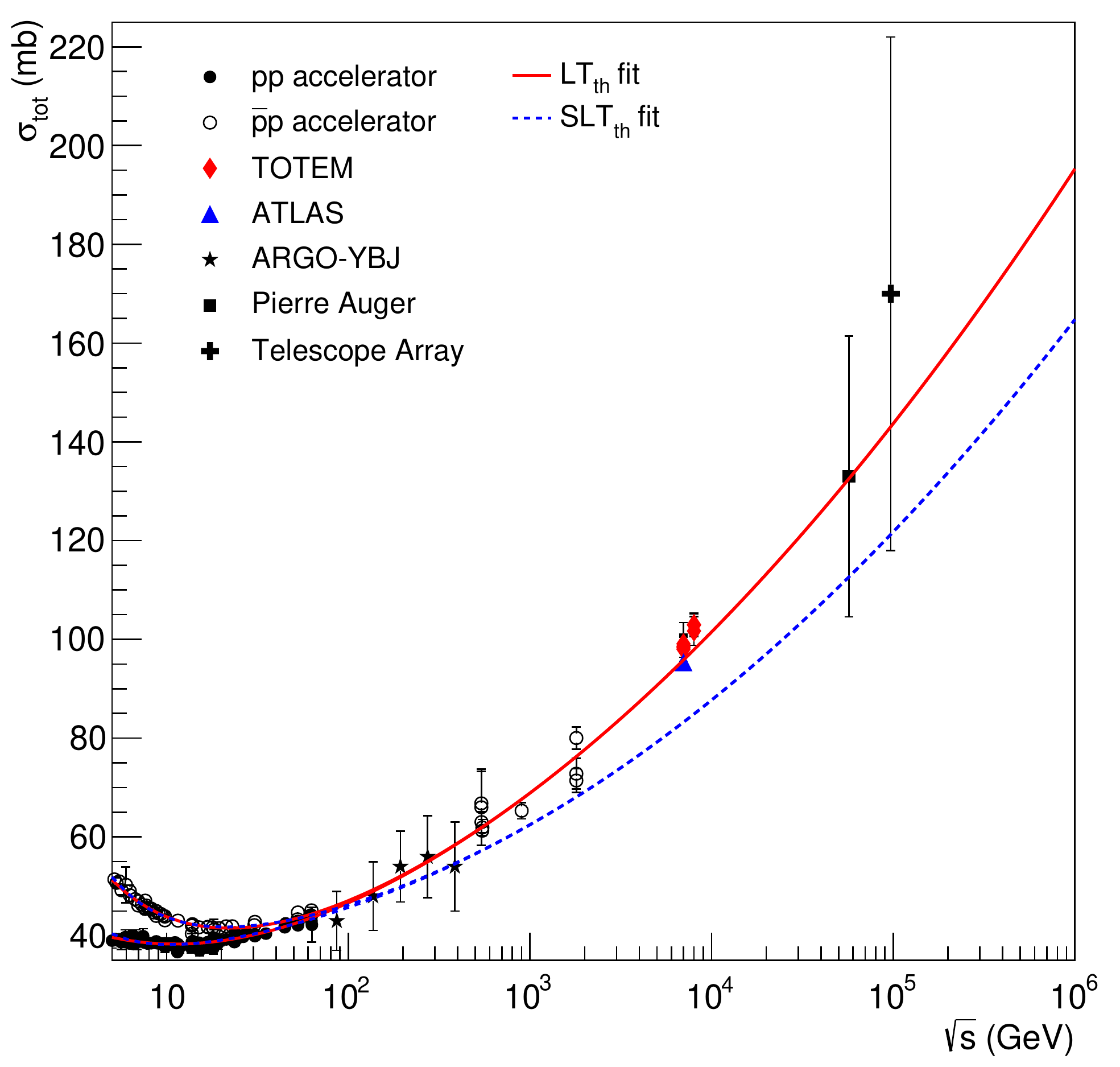}
\caption{\label{fig:res_pp_pbarp_SLTf}Results of fits with $\LTth$
 ($\kappa=1$, $B=B_\text{th}$, $C=0$) and $\SLTth$ ($\kappa=1$,
 $B=B_\text{th}$, $C=C_\text{th}$) to $pp$ and $\ppbar$ data for
 $B_\text{th}$ and $C_\text{th}$ calculated from the $\Omega^\pm$ baryon.}
\end{figure}

The values of $B$ and $C$ calculated with the mass and spin of
the $\Omega^\pm$ baryon [eq.~\eqref{eq:B_C_th_Omega}] are  $B_\text{th} =
0.22$ mb and $C_\text{th} = -0.44$ mb. In $\LTth$ we 
have $B=B_\text{th}$ and $C=0$ fixed. The result obtained (2nd
column of table \ref{tab:res_pp_pbarp_SLTkf}) is close to the one
obtained using $\LT$. This is not surprising, given that in that case,
(see table \ref{tab:res_pp_pbarp}) one finds $B=0.2269$~mb~$\sim$~0.23~mb,
close to $B_\text{th}$, as we have already remarked.
The other parameters and the value of $\chi^2/\nu$ present small
variations compared to $\LT$. We also have a good description of the 
experimental data (see figure \ref{fig:res_pp_pbarp_SLTf}). 

On the other hand, in $\SLTth$, where now $C = C_\text{th}$ is fixed,
the result is not satisfactory as with the previous variant. In fact, the
value of $\chi^2/\nu$ is considerably high, indicating a poor description
of the data. This can be seen in figure \ref{fig:res_pp_pbarp_SLTf}:
the curve obtained from the fit does not describe data with $\sqrt{s}
\gtrsim 1$ TeV. We would like to point out that we 
have practically no change in the parameters associated with the
odd signature Reggeon contribution, while the intercept of the even
trajectory increases going from $\LTth$ to $\SLTth$. The same happens
to $A_1$.
 
\paragraph{Fits with $\LTth$ and $\SLTth$: $B$ and $C$ from the $2^{++}$ glueball state.}
 
In this case, we consider the (quenched) mass and spin of the
$2^{++}$ glueball state, giving $B_\text{th}=-C_\text{th}=0.42$~mb
[eq.~\eqref{eq:B_C_th_g2pp}]. With both variants we get a poor
description of the data with $\chi^2/\nu \sim 3$.
Besides this, the fits present a non-positive-definite error matrix.
In this case, although the fit has converged, we cannot fully trust
the uncertainties estimated for the free parameters~\cite{MINUIT_Manual1994},
and so the results of our fit in general.
For these reasons we have decided not to 
show the results in table \ref{tab:res_pp_pbarp_SLTkf}. For the sake
of completeness, we mention that the intercept $b_1$ of the even Reggeon
trajectory is very small compared with that obtained in the other cases
and that we have a negative ``constant'' Pomeron contribution ($\AP<0$).
 
The description of data are similar for the two variants, with
overestimation of $\sigmatot$ at LHC energies.
  
\paragraph{Fits with $\LTth$ and $\SLTth$: $B$ and $C$ from the $3^{+-}$ glueball state.}
  
Considering the $3^{+-}$ glueball state, we have (using again the
quenched mass) $B_\text{th}=0.78$ mb and $C_\text{th}~=~-~0.39$~mb [see
eq. \eqref{eq:B_C_th_g3pm}]. In these cases, the minimizer did not
converge and, therefore, no fit results were obtained.

\begin{table*}
   \caption{\label{tab:res_pp_pbarp_SLTkf}Results of fits with
     $\LTth$ ($\kappa=1$, $B=B_\text{th}$, $C=0$),
     $\SLTth$ ($\kappa=1$, $B=B_\text{th}$, $C=C_\text{th}$) and
     $\SLTthk$ ($\kappa$ free, $B=B_\text{th}$, $C=C_\text{th}$) to
     $\sigmatot$ data of $\pp$ and $\ppbar$ scattering. 
     The values of $B$ and $C$ are fixed to the
     theoretical values calculated with the masses and the spins of
     the $\Omega^{\pm}$ baryon, the $2^{++}$ glueball state and the
     $3^{+-}$ glueball state (quenched values). For the units of
     measurement of the parameters, see table \ref{tab:res_pp_pbarp}.}
   \begin{tabular}{c|ccc|c|c}\hline\hline
                & \multicolumn{3}{c|}{$\Omega^\pm$ baryon} & {$2^{++}$ glueball} & {$3^{+-}$ glueball} \\\hline
                & $\LTth$ & $\SLTth$ & $\SLTthk$ & $\SLTthk$ & $\SLTthk$ \\\hline
   $B$          & 0.22 (fixed)&  0.22 (fixed) &  0.22 (fixed)  &  0.42 (fixed)  &  0.78 (fixed)      \\
   $C$          & 0 (fixed)   & -0.44 (fixed) & -0.44 (fixed)  & -0.42 (fixed)  & -0.39 (fixed)      \\
   $b_1$        & 0.365(10)   &  0.743(20)    &  0.548(20)     &  0.385(17)     &  0.361(17)         \\
   $b_2$        & 0.539(15)   &  0.528(16)    &  0.540(15)     &  0.539(15)     &  0.539(15)         \\
   $\kappa$     & 1 (fixed)   &  1 (fixed)    &  1.377(18)     &  0.6159(96)    &  0.3097(51)        \\
   $A_1$        & 58.5(1.7)   &  115.3(8.5)   &  57.5(3.2)     &  56.0(2.2)     &  56.3(2.0)         \\
   $A_2$        & 35.3(2.5)   &  33.7(2.4)    &  35.4(2.5)     &  35.3(2.4)     &  35.2(2.4)         \\
   $\AP$        & 25.75(21)   &  35.862(74)   &  32.17(29)     &  28.13(46)     &  26.38(55)         \\\hline
   $\chi^2/\nu$ & 0.987       &  3.59         &  0.937         &  0.957         &  0.965             \\
   $\nu$        & 166         &  166          &  165           &  165           &  165               \\\hline\hline
   \end{tabular}
\end{table*}

\paragraph{Fits with $\SLTthk$ ($\kappa$ free, $B=B_\text{th}$, $C=C_\text{th}$).}

In these variants we consider both $B$ and $C$ fixed with $\kappa$
free. The parameters obtained with these fits are shown in table
\ref{tab:res_pp_pbarp_SLTkf}. With these variants we 
can estimate which of the scenarios discussed after 
eq.~\eqref{eq:ratio_eltot_kappa} is realized at asymptotically high
energies, since according to
eq.~\eqref{eq:ratio_eltot_kappa} the asymptotic ratio between
$\sigmael$ and $\sigmatot$ is given by $\kappa/2$.

\begin{figure}
\centering
\includegraphics[scale=0.45]{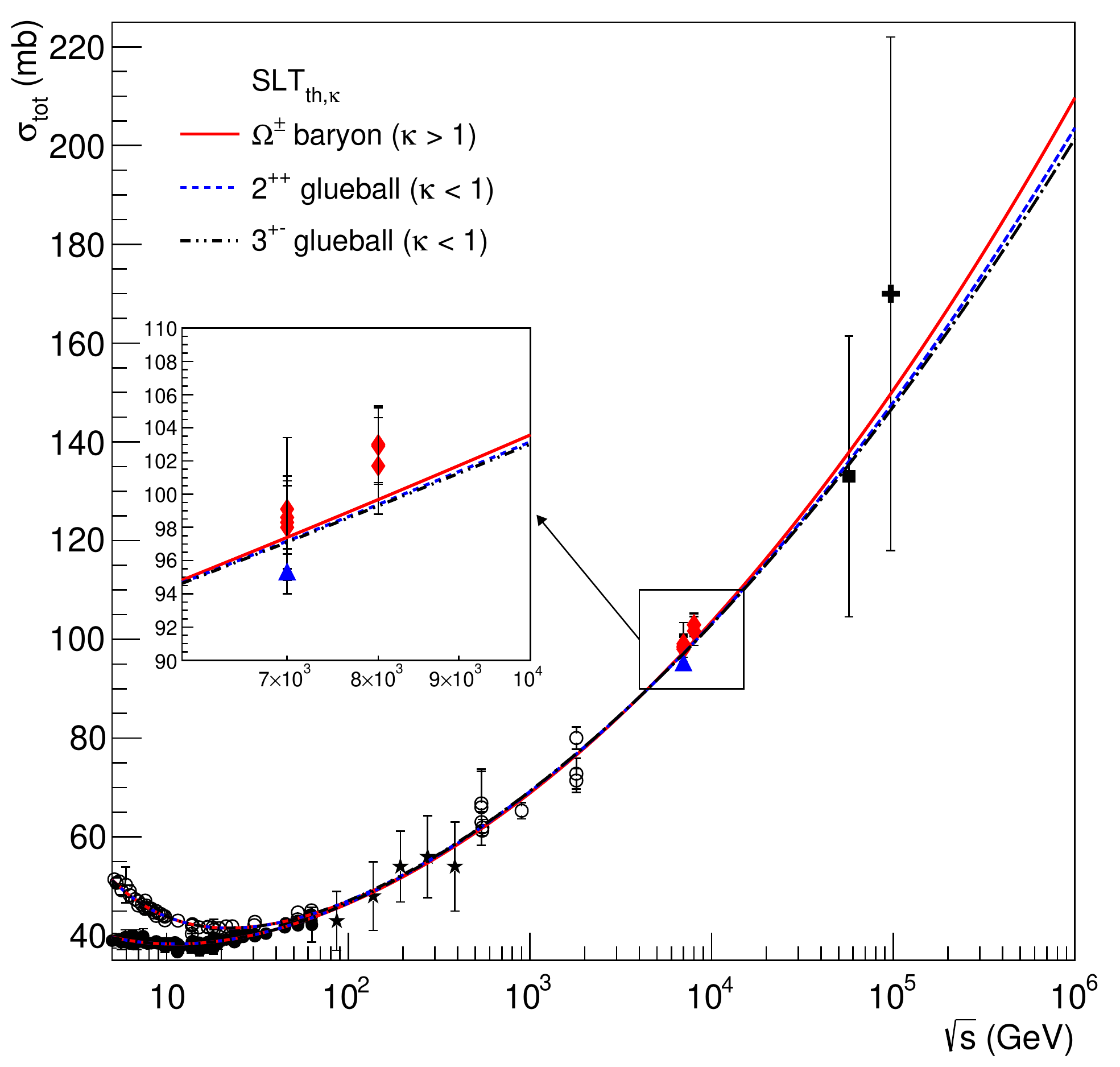}
\caption{\label{fig:res_pp_pbarp_SLTkf}Results of fits with $\SLTthk$
 ($\kappa$ free, $B=B_\text{th}$, $C=C_\text{th}$) to $pp$ and
 $\ppbar$ data for $B_\text{th}$ and $C_\text{th}$ calculated from the
 $\Omega^\pm$ baryon, and the $2^{++}$ and $3^{+-}$ glueball states.
 The legend of data is the same of figure \ref{fig:res_pp_pbarp}.} 
\end{figure}

In all the cases considered for the $B_\text{th}$ and $C_\text{th}$
values, the fits are of good quality ($\chi^2/\nu \lesssim 1$) with
small differences in the $\chi^2$ value among them. We see small
variations of some parameters, for instance, $A_1$, $b_1$ and
$\AP$. Apart from these differences, the
quality of the matching between the fitted curve and the experimental
data is the same in the fitted energy range for all cases. Indeed, as
can be seen in the inset of fig.~\ref{fig:res_pp_pbarp_SLTkf}, the
difference at LHC energies is very small and the results start to be
appreciably different only at cosmic-ray energies. 

Regarding the value of $\kappa$, using the values of $B_\text{th}$ and
$C_\text{th}$ obtained from the $\Omega^\pm$ baryon, we get $\kappa > 1$,
therefore an \textit{antishadowing} scenario.
For the glueball cases, we get $\kappa < 1$, hence a \textit{grey-disk}
scenario, the value obtained with the $2^{++}$ glueball being larger than
that obtained with the $3^{+-}$.

\subsection{Fits to All Reactions}
\label{sec:results_fits_allreactions}

In this section we present and discuss the results obtained from fits
to the full dataset (data from meson-baryon and baryon-baryon
scattering) with the variants $\LT$ and $\SLT$ (table
\ref{tab:res_allreactions_woSmp}) and 
with the variants $\LTth$, $\SLTth$ 
and $\SLTthk$ (table \ref{tab:res_allreactions_SLTkf_woSmp}).
The comparison of the corresponding curves with the experimental data
is shown in figs. \ref{fig:res_allreactions_woSmp},
\ref{fig:res_allreactions_woSmp_SLTf_Omega} and
\ref{fig:res_allreactions_woSmp_SLTkf}.

\subsubsection{Fits with $\LT$ and $\SLT$}
\label{sec:res_allreactions_SLT}

\begin{figure*}
\centering
\includegraphics[scale=0.8]{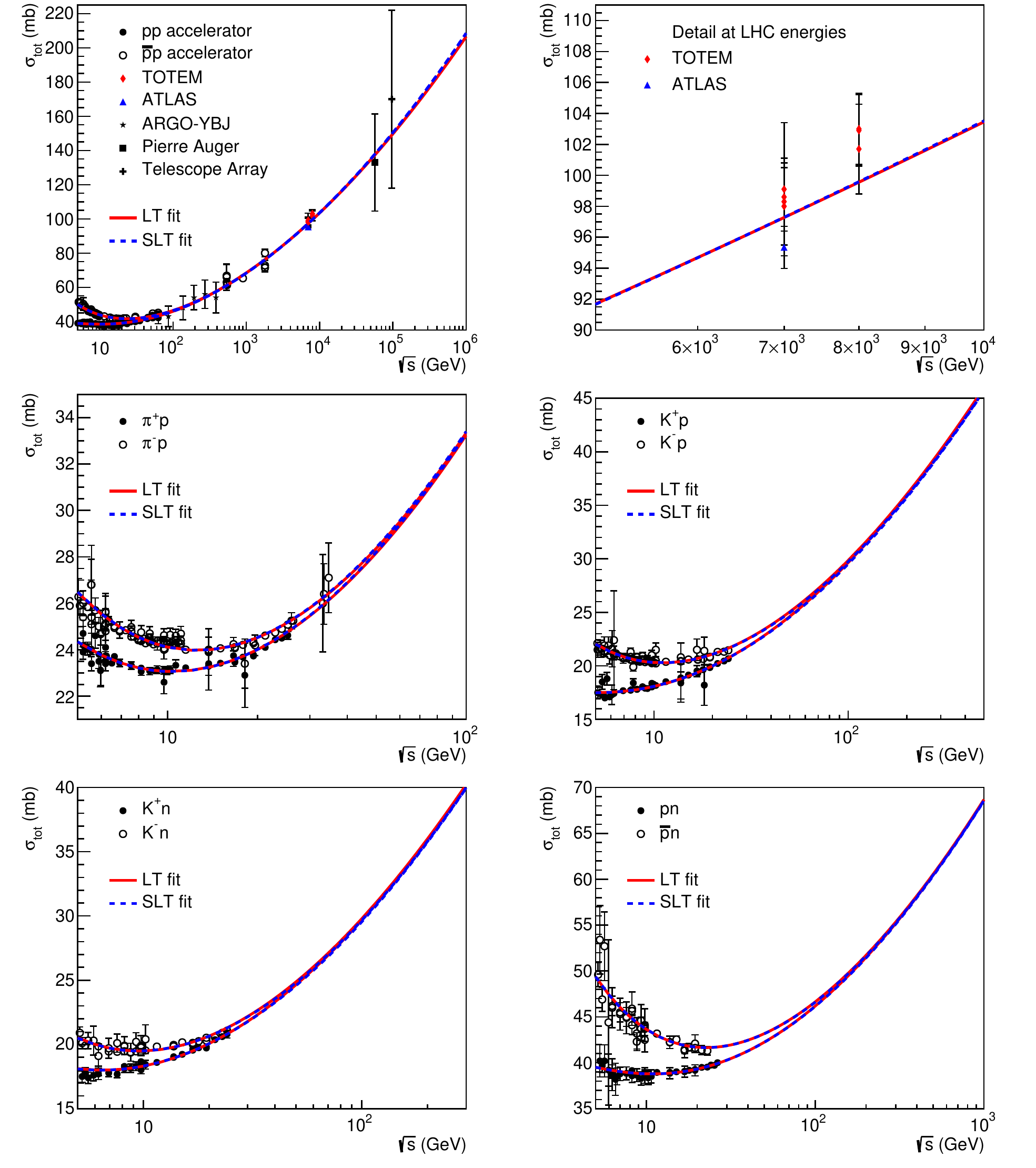}
\caption{\label{fig:res_allreactions_woSmp}Results of fits with $\LT$ and $\SLT$ to all data.}
\end{figure*}

The results are presented in table \ref{tab:res_allreactions_woSmp} and in
figure \ref{fig:res_allreactions_woSmp}. With the $\LT$ variant we obtain
a good description of the data with $\chi^2/\nu\sim 1$.
The same is true for $\SLT$, that we now discuss in some detail. 

As already mentioned above in section \ref{sec:methodology}, the
$\SLT$ fit requires some care. Following the same scheme for the
choice of initial values as in the analysis of $pp$ and $\ppbar$ data (see
fig.~\ref{fig:IV_scheme}), the resulting fit has a
non-positive-definite error matrix. We therefore decided to use the
values obtained for the parameters $B$ and $C$ in the fit
to $\pp$ and $\ppbar$ data with the $\SLT$
(table \ref{tab:res_pp_pbarp}) instead of the values obtained in the
$\LT$ fit to all reactions. Namely, we used as initial values $B=0.349$
mb and $C=-0.95$ mb instead of $B=0.2433$ mb and $C=0$.
For the other parameters the initial values remain unchanged, i.e., 
we used the values obtained in the $\LT$ fit to all data.
With this choice we obtained a more reliable result with an accurate
error matrix.  
In the $\SLT$ fit to all data we find again $C<0$, but with smaller
magnitude and uncertainty than in the $\SLT$ fit to $pp$ and $\ppbar$
data, although the relative uncertainty is the same ($\sim$ 
22\%). We attribute this to the presence of more data at
low energies. On the other hand, the $\chi^2/\nu$ is practically the same.
It is important to mention that here we are increasing the effect of
low-energy data in the estimation of $C$ compared to the $pp$/$\ppbar$
fits, since we have more low-energy  
than high-energy data points in the present dataset. In fact, we have
non-zero correlation coefficients between low- and high-energy parameters,
indicating the influence of the low-energy data in the determination of
$C$ in the fit (see also table 6 in Ref.~\cite{Menon_Silva2013a}).  

Apart from these general aspects of the fits, there is still one point
that demands some comments.
This point concerns the negative value of the parameter $\AP^{\pi p}$
that appears in $\LT$, while it changes to a positive value in $\SLT$.
This is the only negative constant Pomeron contribution (although with large
errors) in this set of fits.
Taking into account the property of factorization of the residues of the
Regge poles~\cite{Barone_Predazzi_book2002} (see also the comments in
Ref.~\cite{COMPETE2002}), this parameter is expected to be positive. However,
factorization is only proven in the case of simple poles and it is valid
when only one trajectory dominates~\cite{Barone_Predazzi_book2002}.
As already commented in footnote \ref{foot:const_term} in
sec.~\ref{sec:parametrization}, $\AP^{ab}$ could also originate from
the sum of a constant (simple-pole) Pomeron with the constant term
coming from higher-order singularities 
(e.g., a triple-pole Pomeron) and, in any case (see also the discussion
at the end of sec.~\ref{sec:parametrization}), its value is affected by the
choice of the energy scale in the leading and subleading terms
in eq. \eqref{eq:sigtot_he}.
Therefore, we cannot exclude this result only on the basis of
factorization. On the other hand, even with $\AP^{\pi p}<0$,
we do not have a negative Pomeron contribution (from the combination
of this constant term plus the $\ln^2 s$ term) in $\LT$.
 
\begin{table}
  \caption{\label{tab:res_allreactions_woSmp}Results of fits with
    $\LT$ and $\SLT$ to $\sigmatot$ (all data).
    For the units of measurement of the parameters, see table
    \ref{tab:res_pp_pbarp}.} 
  \begin{tabular}{c|ccc}\hline\hline
                     & $\LT$           &  $\SLT$     \\\hline
  $B$                & 0.2433(46)      &  0.2652(96) \\ 
  $C$                & 0 (fixed)       & -0.200(44)  \\ 
  $b_1$              & 0.222(11)       &  0.2420(85) \\ 
  $b_2$              & 0.5128(99)      &  0.513(11)  \\ 
  $\kappa$           & 1 (fixed)       &  1 (fixed)  \\\hline
  $A_1^{pp}$         & 47.86(62)       &  44.33(91)  \\ 
  $A_2^{pp}$         & 30.8(1.4)       &  30.8(1.5)  \\ 
  $\AP^{pp}$         & 19.0(1.1)       &  22.61(22)  \\\hline
  $A_1^{pn}$         & 47.2(1.1)       &  43.6(1.3)  \\ 
  $A_2^{pn}$         & 27.4(1.5)       &  27.5(1.6)  \\ 
  $\AP^{pn}$         & 19.2(1.1)       &  22.86(36)  \\\hline
  $A_1^{\pi p}$      & 70.37(99)       &  67.9(1.7)  \\ 
  $A_2^{\pi p}$      & 15.7(1.0)       &  15.8(1.1)  \\ 
  $\AP^{\pi p}$      & -3.3(1.3)       &  0.80(31)   \\\hline
  $A_1^{Kp}$         & 3.42(57)        &  30.31(73)  \\ 
  $A_2^{Kp}$         & 17.54(91)       &  17.56(96)  \\ 
  $\AP^{Kp}$         & 1.77(85)        &  5.09(11)   \\\hline
  $A_1^{Kn}$         & 32.72(73)       &  28.76(77)  \\ 
  $A_2^{Kn}$         & 9.28(69)        &  9.30(71)   \\ 
  $\AP^{Kn}$         & 1.93(84)        &  5.22(14)   \\\hline
  $\chi^2/\nu$       & 1.060           &  1.063      \\
  $\nu$              & 532             &  531        \\\hline\hline
  \end{tabular}
\end{table} 
 
\subsubsection{Fits with $\LTth$ ($\kappa=1$, $B=B_\text{th}$, $C=0$),
 $\SLTth$ ($\kappa=1$, $B=B_\text{th}$, $C=C_\text{th}$) and
 $\SLTthk$ ($\kappa$ free, $B=B_\text{th}$, $C=C_\text{th}$)}
\label{sec:res_allreactions_SLTf}
 
In table \ref{tab:res_allreactions_SLTkf_woSmp} we present the results obtained
with the variants $\LTth$ and $\SLTth$, where $B$ and $C$ are fixed to their
theoretical values [see eqs.~\eqref{eq:B_C_th_Omega}--\eqref{eq:B_C_th_g3pm}].
 
\paragraph{Fits with $\LTth$ and $\SLTth$: $B$ and $C$ from the $\Omega^\pm$ baryon.}
 
The results (2nd and 3rd column of table
\ref{tab:res_allreactions_SLTkf_woSmp}) obtained in this case are 
satisfactory for $\LTth$ with only a small increase of $\chi^2/\nu$
in comparison with $\LT$.
For $\SLTth$, we get $\chi^2/\nu\sim 2$ while in $\SLT$ we have $\sim 1$.
However, this increase is less than that observed in the fits to $pp$ and
$\ppbar$ data only. Contrary to the $\LT$ fit, we have that all $\AP^i>0$
in $\LTth$.
  
\begin{figure*}
\centering
\includegraphics[scale=0.8]{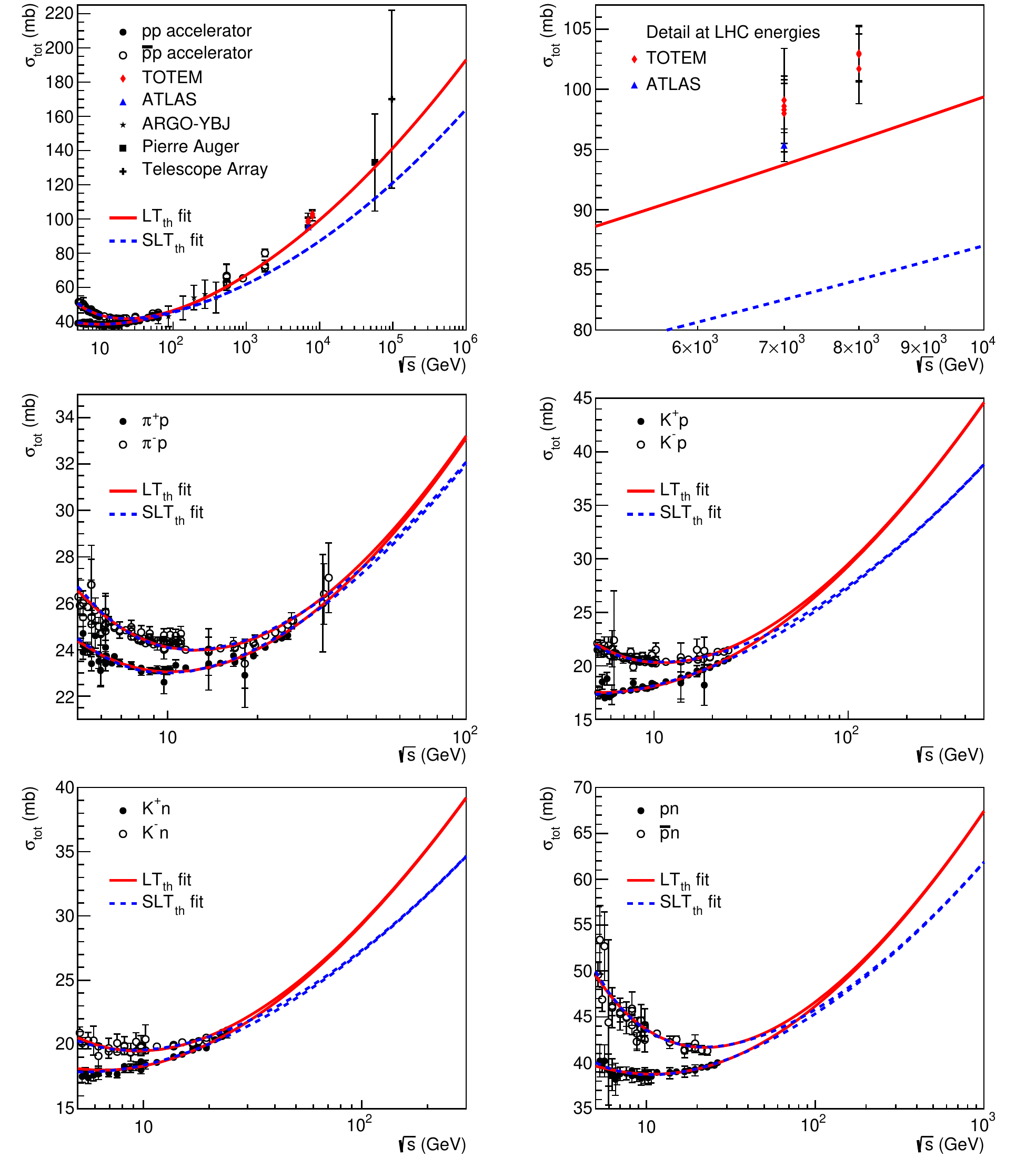}
\caption{\label{fig:res_allreactions_woSmp_SLTf_Omega}Results of fits
 with $\LTth$ ($\kappa=1$, $B=B_\text{th}$, $C=0$) and $\SLTth$
 ($\kappa=1$, $B=B_\text{th}$, $C=C_\text{th}$) to all data for
 $B_\text{th}$ and $C_\text{th}$ calculated from the $\Omega^\pm$ baryon.}
\end{figure*}

Regarding the comparison with the data, $\LTth$ gives a good
description of them, favoring the ATLAS point at 7 TeV in $pp/\ppbar$,
while $\SLTth$ does not seem to describe well the $pp/\ppbar$ data with
$\sqrt{s} \gtrsim 500$ GeV (see
fig.~\ref{fig:res_allreactions_woSmp_SLTf_Omega}). 
Concerning the other reactions (where data are at much lower energies than in
the $pp/\ppbar$ case), both $\LTth$ and $\SLTth$ seem to describe data
reasonably well (even if, of course, the two variants have a quite different
high-energy behavior).
 
\paragraph{Fits with $\LTth$ and $\SLTth$: $B$ and $C$ from the $2^{++}$ glueball state.}
 
The fits with $\LTth$ and $\SLTth$ in this case have a non-positive-definite
error matrix. The results (not presented in table
\ref{tab:res_allreactions_SLTkf_woSmp}) have the
same features of the fits to $pp$ and $\ppbar$ data, for example, a
very small $b_1$ parameter. We also have that almost all $\AP^i <0$. The
$\chi^2/\nu$ values are around 1.5, with a similar quality in the
matching with data for both variants.
Regarding $pp$ and $\ppbar$, the fits overestimate the data at LHC energies,
reaching the upper error bar of the TOTEM data.
  
\paragraph{Fits with $\LTth$ and $\SLTth$: $B$ and $C$ from the $3^{+-}$ glueball state.}
 
Again, using the mass and spin of the $3^{+-}$ glueball state, the
fits did not converge.  
 
\begin{table*}
  \caption{\label{tab:res_allreactions_SLTkf_woSmp}Results of fits with
    $\LTth$ ($\kappa=1$, $B=B_\text{th}$, $C=0$),
    $\SLTth$ ($\kappa=1$, $B=B_\text{th}$, $C=C_\text{th}$) and
    $\SLTthk$ ($\kappa$ free, $B=B_\text{th}$, $C=C_\text{th}$) to
    $\sigmatot$ (all data). The values of
    $B$ and $C$ are fixed to the theoretical values calculated with
    the masses and the spins of the $\Omega^{\pm}$ baryon, the
    $2^{++}$ glueball state and the $3^{+-}$ glueball state (quenched
    values). For the units of measurement of the parameters, see table
    \ref{tab:res_pp_pbarp}.}
  \begin{tabular}{c|ccc|c|c}\hline\hline
                     & \multicolumn{3}{c|}{$\Omega^\pm$ baryon}& {$2^{++}$ glueball} & {$3^{+-}$ glueball}      \\\hline
                     & $\LTth$      & $\SLTth$      & $\SLTthk$     & $\SLTthk$     & $\SLTthk$       \\\hline
  $B$                & 0.22 (fixed) &  0.22 (fixed) &  0.22 (fixed) &  0.42 (fixed) &  0.78 (fixed)  \\ 
  $C$                & 0 (fixed)    & -0.44 (fixed) & -0.44 (fixed) & -0.42 (fixed) & -0.39 (fixed)  \\ 
  $b_1$              & 0.2744(66)   &  0.554(13)    &  0.292(14)    &  0.249(13)    &  0.234(12)     \\ 
  $b_2$              & 0.5141(97)   &  0.515(11)    &  0.514(10)    &  0.513(11)    &  0.513(11)     \\ 
  $\kappa$           & 1 (fixed)    &  1 (fixed)    &  1.439(23)    &  0.653(12)    &  0.3303(64)    \\\hline 
  $A_1^{pp}$         & 47.04(71)    &  59.0(2.7)    &  37.99(87)    &  43.12(57)    &  45.54(58)     \\ 
  $A_2^{pp}$         & 31.0(1.4)    &  31.4(1.6)    &  30.9(1.4)    &  30.8(1.5)    &  30.8(1.5)     \\ 
  $\AP^{pp}$         & 23.40(24)    &  35.159(77)   &  29.22(51)    &  23.76(82)    &  21.29(98)     \\\hline 
  $A_1^{pn}$         & 46.3(1.2)    &  57.8(3.2)    &  37.3(1.2)    &  42.4(1.1)    &  44.9(1.1)     \\ 
  $A_2^{pn}$         & 27.6(1.5)    &  27.9(1.7)    &  27.5(1.6)    &  27.5(1.6)    &  27.4(1.6)     \\ 
  $\AP^{pn}$         & 23.64(37)    &  35.24(15)    &  29.44(56)    &  24.01(87)    &  21.5(1.0)     \\\hline 
  $A_1^{\pi p}$      & 73.5(1.5)    &  136.9(8.2)   &  64.8(2.3)    &  67.0(1.5)    &  68.6(1.2)     \\ 
  $A_2^{\pi p}$      & 16.11(98)    &  16.7(1.1)    &  16.1(1.0)    &  15.9(1.1)    &  15.8(1.1)     \\ 
  $\AP^{\pi p}$      & 19.84(28)    &  15.921(83)   &  8.27(65)     &  20.8(1.0)    & -0.69(1.2)     \\\hline
  $A_1^{Kp}$         & 32.53(51)    &  28.1(1.4)    &  22.61(49)    &  28.95(47)    &  31.68(54)     \\ 
  $A_2^{Kp}$         & 17.67(88)    &  17.66(98)    &  17.57(92)    &  17.55(96)    &  17.54(97)     \\ 
  $\AP^{Kp}$         & 5.39(18)     &  15.450(48)   &  11.09(37)    &  6.15(64)     &  3.90(78)      \\\hline
  $A_1^{Kn}$         & 30.89(72)    &  23.8(1.8)    &  20.86(64)    &  27.38(71)    &  30.16(72)     \\
  $A_2^{Kn}$         & 9.35(69)     &  9.43(73)     &  9.33(70)     &  9.30(71)     &  9.29(72)      \\ 
  $\AP^{Kn}$         & 5.48(20)     &  15.388(71)   &  11.18(36)    &  6.27(64)     &  4.04(77)      \\\hline 
  $\chi^2/\nu$       & 1.108        &  1.966        &  1.071        &  1.062        &  1.061         \\
  $\nu$              & 533          &  533          &  532          &  532          &  532           \\\hline\hline
  \end{tabular}
\end{table*}
 
\paragraph{Fits with $\SLTthk$ ($\kappa$ free, $B=B_\text{th}$, $C=C_\text{th}$).}

\begin{figure*}[h!]
\centering
\includegraphics[scale=0.8]{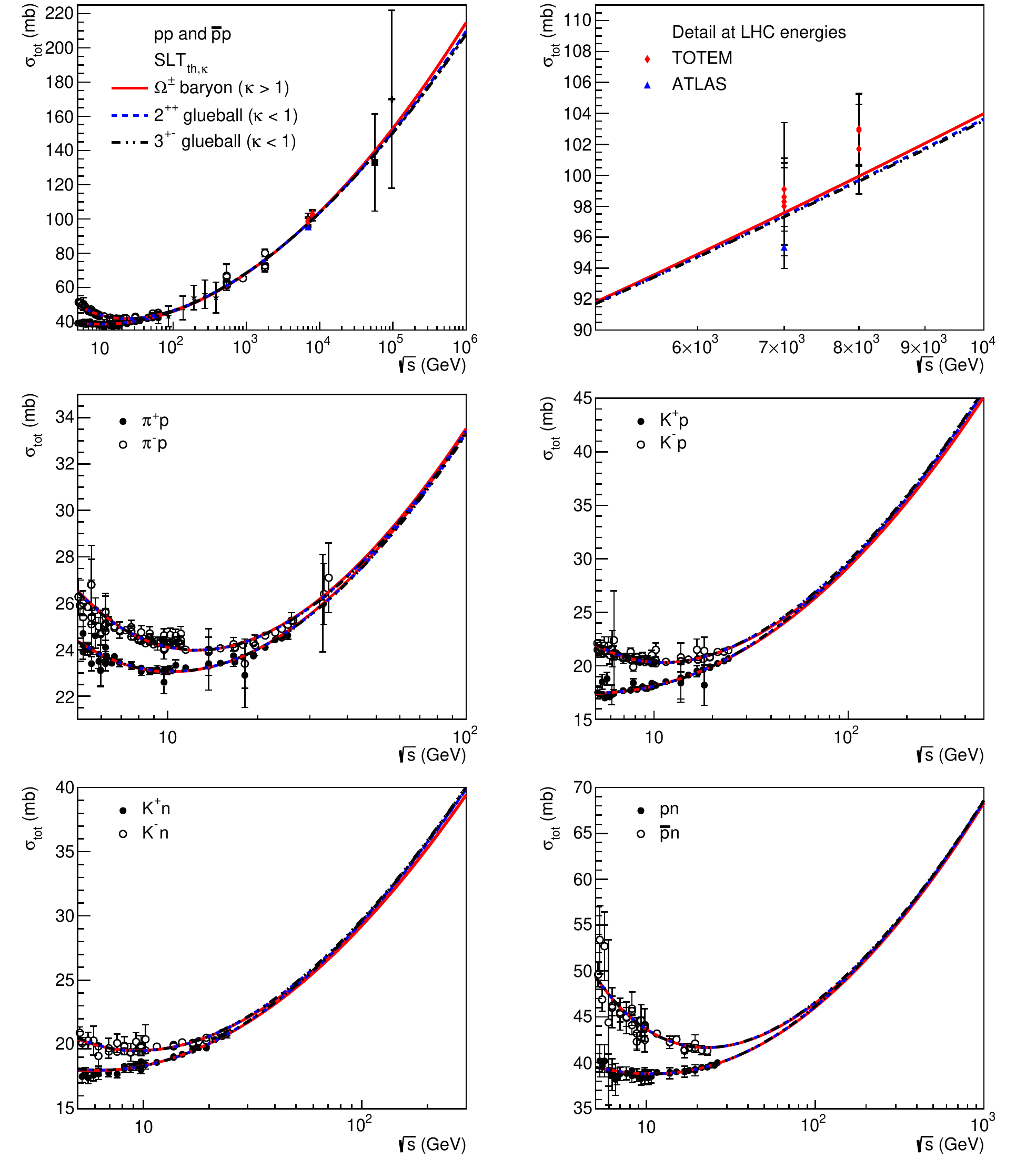}
\caption{\label{fig:res_allreactions_woSmp_SLTkf}Results of fits with
 $\SLTthk$ ($\kappa$ free, $B=B_\text{th}$, $C=C_\text{th}$) to all
 data for $B_\text{th}$ and $C_\text{th}$ calculated from the
 $\Omega^\pm$ baryon, and the $2^{++}$ and $3^{+-}$ glueball states.
 The legend for the curves is shown in the top-left panel. For the
 legend of $pp$ and $\ppbar$ data see figure \ref{fig:res_pp_pbarp}.}
\end{figure*}

Finally we discuss the results of the fits where $\kappa$ is a free
parameter. We show in table \ref{tab:res_allreactions_SLTkf_woSmp} all
the parameters determined in the fits.
The $\chi^2/\nu$ values are close to 1, indicating that the fit is of
good quality. Furthermore, we get a good description of
data in all cases.  As in the fit with $\SLT$, the ``constant''
Pomeron term $\AP^{\pi p}$ is positive, except when we consider the
$3^{+-}$ glueball state, in which case we find a negative central
value, which is however compatible with zero. 

The agreement between the fitted curve and the data
(fig.~\ref{fig:res_allreactions_woSmp_SLTkf}) is also similar to that
achieved when fitting only the $pp$ and $\ppbar$ data: in the 
energy range of the fit all cases give almost indistinguishable
curves, which present small differences in the extrapolation to higher
energies. In conclusion, the three choices for $B_\text{th}$ and
$C_\text{th}$ all give equally good fits. 

Concerning the parameter $\kappa$, we have $\kappa > 1$ for the $\Omega^{\pm}$
baryon and $\kappa < 1$ for the $2^{++}$ and $3^{+-}$ glueball states
(with smaller $\kappa$ in the latter case),
as it was also found when fitting $pp$ and $\ppbar$ data only,
but with slightly bigger central values.
 
 
\section{Conclusions}
\label{sec:conclusions}

In this work we have performed a phenomenological analysis of total
cross section data from hadronic scattering in order to estimate the 
relevance of a subleading $\ln (s/s_0)\ln\ln(s/s_0)$ term, obtained in
recent theoretical studies in nonperturbative
QCD~\cite{Giordano_Meggiolaro2014} (see also
Refs.~\cite{Martin_Roy2014,Nastase_Sonnenschein2015,Diez_etal2015} for
a similar subleading term). More precisely, we used the following
parametrization of the total cross section in the high-energy region:
\begin{equation}
 \sigmatot^{ab}(s)\! \underset{s\to\infty}{\sim}\! B
 \ln^2\left(\frac{s}{s_0^{ab}}\right) + C
 \ln\left(\frac{s}{s_0^{ab}}\right)
 \ln\left[\ln\left(\frac{s}{s_0^{ab}}\right)\right]\,,
 \label{eq:sigtot_asymp_concl}
\end{equation}
with $s_0^{ab} = m_am_b$. The low-energy region was parametrized as
usual in terms of Reggeon exchange in the $t$-channel.

We have determined the parameters $B$ and $C$ through fits ($\SLT$
fits) to two datasets with total center-of-mass energy $\sqrt{s} \geq$
5 GeV: (i) only $\pp$ and $\ppbar$ scattering and (ii) all
meson-baryon and baryon-baryon data available at the PDG
website~\cite{PDG_data_website} (except for $\Sigma^-p$). In both
cases, the recent data for $pp$ scattering obtained at 7 and 8 TeV 
by the TOTEM and ATLAS
Collaborations~\cite{TOTEM2011b,TOTEM2013a,TOTEM2013c,TOTEM2013d,TOTEM:2016,ATLAS2014} were included in the fits.
In the first case we have obtained $B_{pp} = 0.349 \pm 0.029$ mb
and $C_{pp} = -0.95 \pm 0.21$ mb, while in the second case
$B_\text{all} = 0.2652 \pm 0.0096$ mb and $C_\text{all}=-0.200\pm 0.044$ mb.
Both fits result in a good and equivalent description of the experimental data.
However, we note that $B_{pp}$, $C_{pp}$ and $B_\text{all}$, $C_\text{all}$
are not compatible. We can understand this difference from the fact that
when all reactions are considered we have more data in the low-energy region,
since most of the data of meson-baryon and baryon-baryon scattering are
available for $\sqrt{s} \lesssim 25$ GeV, and this strongly affects the central
value. In any case, it is important to keep in mind that the result
reported in eq. \eqref{eq:sigtot_asymp_concl} is asymptotic and that the
contribution of this subleading term may still be weak, and so
difficult to identify, in the LHC energy region.

As discussed in section \ref{sec:theory}, the parameters $B$ and $C$
may be written, following the nonperturbative QCD approach of
Ref.~\cite{Giordano_Meggiolaro2014},
in terms of the spin $\tilde{s}$ and the mass $\tilde{m}$
of the particle that maximizes the ratio of eq.~\eqref{eq:ratio_s_m}.
This yields (apart from a common multiplicative constant $\kappa$)
what we have called $B_\text{th}$ and $C_\text{th}$ [eq.~\eqref{eq:B_C_th}].
While in Ref.~\cite{Giordano_Meggiolaro2014} $B_\text{th}$ and
$C_\text{th}$ were estimated from the hadronic spectrum, here we can
try to reverse the approach, and estimate $\tilde{s}$ and
$\tilde{m}$ from the values of $B$ and $C$ obtained with the $\SLT$ fit.
Equating the values of $B$ and $C$ in eq.~\eqref{eq:sigtot_asymp_concl}
to $\kappa B_\text{th}$ and $\kappa C_\text{th}$, respectively,
we find that:  
\begin{equation}
   \tilde{s} = 1-\frac{B}{C} \quad \text{and} \quad
   \frac{\tilde{m}^2}{\kappa} = 2\pi \frac{B}{C^2}. 
 \label{eq:ratio_BC_bis}
\end{equation}
Of course, we are not able to determine all the three parameters
$\kappa$, $\tilde{s}$ and $\tilde{m}$ from the knowledge of $B$ and $C$ alone.
This lack of information does not concern the
spin (which can be exactly determined from the ratio $B/C$), but only
the ratio $\tilde{m}/\sqrt{\kappa}$ (instead of the mass $\tilde{m}$
and the parameter $\kappa$ separately) can be determined. 
Using the information of tables \ref{tab:res_pp_pbarp} and
\ref{tab:res_allreactions_woSmp}, 
we have calculated the ratios $B/C$ and
$B/C^2$, from which we have derived $\tilde{s}$ and
$\tilde{m}/\sqrt{\kappa}$. 
With $\SLT$ ($pp/\bar{p}p$) we find $\tilde{s} = 1.367(87)$, which is not far
from the spin $\tilde{s} = 3/2$ of the $\Omega^\pm$ baryon, and
$\tilde{m}/\sqrt{\kappa} = 0.98(21)$ GeV, which for the mass of the
$\Omega^\pm$ baryon implies $\kappa \simeq 2.9 > 2$, i.e., outside of
the unitarity window ($0\le\kappa\le 2$).
Instead, for $\SLT$ (all reactions) we find $\tilde{s} = 2.33(30)$, which
is close to the spin $\tilde{s} = 2$ of the $2^{++}$ glueball, and
$\tilde{m}/\sqrt{\kappa} = 4.03(89)$ GeV, which for the mass of the
$2^{++}$ glueball implies $\kappa \simeq 0.35 < 1$, not far from the
estimate that can be obtained from experimental data (see below).

(For comparison, we recall here two other theoretical predictions for the
ratio $B/C$, obtained using completely different approaches:
in Ref. \cite{Nastase_Sonnenschein2015} the value $B/C = -1/2$ is found,
which agrees with the first eq. \eqref{eq:ratio_BC_bis} when $\tilde{s} = 3/2$,
while in Ref. \cite{Diez_etal2015} the value $B/C = -1/4$ is derived.)

We have also performed fits with $B=B_\text{th}$, $C=0$ fixed ($\LTth$) and
$B=B_\text{th}$, $C=C_\text{th}$ fixed ($\SLTth$), 
using for $B_\text{th}$ and $C_\text{th}$ the values
discussed in Ref.~\cite{Giordano_Meggiolaro2014} 
(and recalled in sec.~\ref{sec:theory}), corresponding to the
$\Omega^\pm$ baryon and the $2^{++}$ and $3^{+-}$ glueball states.
In this class of variants ($B$ and $C$ fixed to theoretical values),
the best result is $\LTth$ with $\Omega^\pm$ baryon values (both in
the $pp/\bar{p}p$ and all reactions cases). 
 
 \begin{table}[t]
 \caption{\label{tab:ratio_eltot_pp_ppbar_allreactions}Ratio
   $\sigmael/\sigmatot = \kappa/2$, with $\kappa$ determined from the fit $\LT$
   (considering $B=\kappa B_\text{th}$, i.e., $\kappa = B/B_\text{th}$) and
   from the fit $\SLTthk$ to $pp$ and $\bar{p}p$ data only, and also from fits
   where all data are considered. Uncertainties are calculated with
   standard error propagation.}
 \begin{tabular}{c|cc|cc}\hline\hline
                     &\multicolumn{2}{c|}{Fits to $pp/\bar{p}p$ data only} & \multicolumn{2}{c}{Fits to all data}\\\hline
                     & $\LT$  & $\SLTthk$  & $\LT$ & $\SLTthk$ \\\hline
 $\Omega^\pm$ baryon & 0.5157(86)   & 0.6885(91)   & 0.553(10)    & 0.720(12)     \\
 $2^{++}$ glueball   & 0.2701(45)   & 0.3080(48)   & 0.2896(55)   & 0.3265(60)    \\
 $3^{+-}$ glueball   & 0.1454(24)   & 0.1548(26)   & 0.1560(29)   & 0.1652(32)    \\\hline\hline
 \end{tabular}
\end{table}

Finally, we have also considered fits where the $B$ and $C$ parameters 
are set to $B = \kappa B_\text{th}$ and $C = \kappa C_\text{th}$,
with $\kappa$ treated as an extra free parameter, which is associated to the
asymptotic value of the ratio $\sigmael/\sigmatot$
[eq.~\eqref{eq:ratio_eltot_kappa}]. 
In table \ref{tab:ratio_eltot_pp_ppbar_allreactions} we display all the
resulting values together with the corresponding uncertainty (calculated via
standard error propagation). The fits with the $\Omega^\pm$ baryon values
indicate an asymptotic scenario in the anti-shadowing regime, since
$\sigmael/\sigmatot > 0.5$. 
On the other hand, the values inferred from the fits with $B$ and $C$
fixed to the values obtained from the $2^{++}$ glueball state indicate a
\textit{grey-disk} scenario, with $\sigmael/\sigmatot \sim 0.3$.  
This value is in agreement with the asymptotic ratio obtained in other studies:

\begin{itemize}
 \item empirical fits to the $\sigmael/\sigmatot$ data made by
   Fagundes, Menon and Silva~\cite{Fagundes_Menon_Silva2016a};
 
 \item independent fits to $\sigmatot$ and $\sigmael$ data also made
   by Fagundes, Menon and
   Silva~\cite{Fagundes_Menon_Silva2013a,Menon_Silva2013a,Menon_Silva2013b}
   that yield this ratio close to 0.3;
 
 \item the prediction made by Kohara, Ferreira, and
   Kodama~\cite{Kohara_Ferreira_Kodama2014}, using the Stochastic Vacuum Model
   and fits to elastic scattering data, that this ratio is below 1/2 (and
   close to 1/3);

 \item the scenario of a \textit{black torus} proposed by
   Dremin~\cite{Dremin2014,Dremin2015a,Dremin2015b}, where the particle has a
   semi-transparent center (\textit{grey disk}) surrounded by a
   \textit{black disk}. This scenario has also been proposed, in a different
   context, in Refs.~\cite{Desgrolard_Jenkovszky_Struminsky1999,Desgrolard_Jenkovszky_Struminsky2000}.
\end{itemize}

Using the mass and spin of the $3^{+-}$ glueball state, we also get a
\textit{grey-disk} scenario: however, the resulting asymptotic value is smaller
than the experimental data available so far. From table
\ref{tab:ratio_eltot_pp_ppbar_allreactions}, we see that the value for
this ratio is around 0.15, while the experimental value at the highest
energy obtained so far (8 TeV) is approximately 0.27. 
Since by now the data show a rising trend with energy (see 
for instance fig.~1 in Ref.~\cite{Fagundes_Menon_Silva2016a}),
this means that if this scenario is the correct one, then the data must
present a local maximum and then decrease as the energy increases
until it reaches the asymptotic value. Although there seems to be
no theoretical reason to exclude this type of behavior, it seems
quite unlikely to happen, and we would rather expect a smooth
rise with energy until the asymptotic value
is reached.

We recall that, as mentioned in section \ref{sec:theory}, the
result obtained in Ref.~\cite{Giordano_Meggiolaro2014} also includes a
subleading term $Q^{ab}\ln s$, where $Q^{ab}$ is a reaction-dependent
parameter. We avoided this term in the present analysis in order to
study the universality of the leading ($\ln^2 s$) and subleading ($\ln
s \ln\ln s$) terms, leaving it to future work. We also remark that the
universality of $\kappa$ is part of our assumptions.

Another important aspect to keep in mind is that $\ln s \ln\ln s$ is a
slowly varying function, so that we may need high-energy data in a
larger range than what is currently available in order to be
able to properly estimate the contribution of the subleading term. 
Therefore, new measurements of $\sigmatot$ (and also $\sigmael$)
at 13 TeV and higher energies are fundamental
to improve and confirm the results presented here.  

\begin{acknowledgments}
 P.V.R.G. Silva thanks the financial support by 
 S\~ao Paulo Research Foundation (FAPESP) under the
 contracts 2013/27060-3 and 2015/21855-0.
 The authors thanks M.J. Menon for useful discussions.
\end{acknowledgments}

\bibliographystyle{apsrev4-1}
\bibliography{references.bib}

\end{document}